\documentclass[preprintnumbers,article,amsmath,amssymb,floatfix,10pt,prd,twocolumn,
superscriptaddress,nofootinbib]{revtex4-2}
\usepackage{bm}
\usepackage{amsfonts}
\usepackage{latexsym}
\usepackage[latin1]{inputenc}
\usepackage{graphicx}
\usepackage{amsmath}
 \usepackage{palatino}
\usepackage{mathpazo}
\usepackage{textcomp}
\usepackage{float}
\usepackage{booktabs}
\usepackage{dcolumn}
\usepackage{ragged2e}
\usepackage{hyperref}
\hypersetup{colorlinks,citecolor=blue}
\hypersetup{colorlinks=true,linkcolor=red,filecolor=magenta,    urlcolor=blue}
\usepackage{amsmath}
\usepackage{xcolor}
\usepackage{orcidlink}
\usepackage{epsfig}
\usepackage{subfigure}
\usepackage{commath}

\def\jnl@style{\it}
\def\aaref@jnl#1{{\jnl@style#1}}

\def\aaref@jnl#1{{\jnl@style#1}}

\def\aj{\aaref@jnl{AJ}}                   
\def\apj{\aaref@jnl{ApJ}}                 
\def\apjl{\aaref@jnl{ApJ}}                
\def\apjs{\aaref@jnl{ApJS}}               
\def\apss{\aaref@jnl{Ap\&SS}}             
\def\aap{\aaref@jnl{A\&A}}                
\def\aapr{\aaref@jnl{A\&A~Rev.}}          
\def\aaps{\aaref@jnl{A\&AS}}              
\def\mnras{\aaref@jnl{Mon.~Not.~Roy.~Astron.~Soc.}}             
\def\prd{\aaref@jnl{Phys.~Rev.~D}}        
\def\prc{\aaref@jnl{Phys.~Rev.~C}}  
\def\prl{\aaref@jnl{Phys.~Rev.~Lett.}}    
\def\qjras{\aaref@jnl{QJRAS}}             
\def\skytel{\aaref@jnl{S\&T}}             
\def\ssr{\aaref@jnl{Space~Sci.~Rev.}}     
\def\zap{\aaref@jnl{ZAp}}                 
\def\nat{\aaref@jnl{Nature}}              
\def\aplett{\aaref@jnl{Astrophys.~Lett.}} 
\def\apspr{\aaref@jnl{Astrophys.~Space~Phys.~Res.}} 
\def\physrep{\aaref@jnl{Phys.~Rep.}}      
\def\physscr{\aaref@jnl{Phys.~Scr}}       
\def\commat{\aaref@jnl{Comm.~Math.~Phys.}}              
\def\science{\aaref@jnl{Science}}               
\def\cqg{\aaref@jnl{Classical Quant.~Grav.}}            
\def\jpcs{\aaref@jnl{JPCS}}                                     
\def\ijmpd{\aaref@jnl{Int.~J.~Mod.~Phys.~D}}                    
\def\grg{\aaref@jnl{Gen.~Relat.~Gravit.}}               
\def\rpp{\aaref@jnl{Rep.~Prog.~Phys.}}          
\def\npa{\aaref@jnl{Nucl.~Phys.~A}}        
\def\lrr{\aaref@jnl{Living Rev.~Rel.}}                   
\def\jcap{\aaref@jnl{J.~Cosmology Astropart.~Phys.}}    
\def\rmp{\aaref@jnl{Rev.~Mod.~Phys.}}   
\def\epjc{\aaref@jnl{Eur.~Phys.~J.~C}} 
\def\plb{\aaref@jnl{~Phy.~Lett.~B}} 
\def\mpla{\aaref@jnl{Mod.~Phy.~Lett.~A}} 
\def\arxiv{\aaref@jnl{arxiv.org}}

\allowdisplaybreaks[1]

\begin{document}
\color{black}       
\title{GUP Corrected Casimir Wormholes in $f(Q)$ Gravity}

\author{Zinnat Hassan\orcidlink{0000-0002-6608-2075}}
\email{zinnathassan980@gmail.com}
\affiliation{Department of Mathematics, Birla Institute of Technology and Science-Pilani,\\ Hyderabad Campus, Hyderabad-500078, India.}

\author{Sayantan Ghosh\orcidlink{0000-0002-3875-0849}}
\email{sayantanghosh.000@gmail.com}
\affiliation{Department of Mathematics, Birla Institute of Technology and Science-Pilani,\\ Hyderabad Campus, Hyderabad-500078, India.}

\author{P.K. Sahoo\orcidlink{0000-0003-2130-8832}}
\email{pksahoo@hyderabad.bits-pilani.ac.in}
\affiliation{Department of Mathematics, Birla Institute of Technology and Science-Pilani,\\ Hyderabad Campus, Hyderabad-500078, India.}

\author{V. Sree Hari Rao}
\email{vshrao@researchfoundation.in; vshrao@gmail.com}
\affiliation{Foundation for Scientific Research and Technological Innovation, Hyderabad - 500102, India.}
%
\date{\today}
\begin{abstract}
We have systematically presented the effect of the Generalized Uncertainty Principle (GUP) in Casimir wormhole space-time in the recently proposed modified gravity, the so-called symmetric teleparallel gravity, or $f(Q)$ gravity. We consider two famous GUP models, such as the Kempf, Mangano, and Mann (KMM) model and the Detournay, Gabriel, and Spindel (DGS) model, in this study. Also, to find the solutions, we assumed two different $f(Q)$ forms and obtained analytic as well as numerical solutions under the effect of GUP. Besides this, we investigate the solutions with three different redshift functions under an anisotropic fluid located at the throat. Further, we analyzed the obtained wormhole solutions with energy conditions, especially null energy conditions (NEC) at the wormhole's throat, and encountered that some arbitrary quantity disrespects the classical energy conditions at the wormhole throat of radius $r_0$. Later, the ADM mass and the volume integral quantifier are also discussed to calculate the amount of exotic matter required near the wormhole throat. Additionally, we show the behavior of the equation of state parameters under the effect of GUP.
\end{abstract}

\maketitle


\textbf{Keywords:} Casimir wormhole, Generalized Uncertainty Principle (GUP), energy conditions, $f(Q)$ gravity.
\section{Introduction}\label{sec1}
Wormholes and Black holes are the two most fascinating solutions to the field equations of Einstein's General Relativity (GR). The existence of Black holes has been investigated in \cite{abbott1,Abbott2,Akiyama}, whereas the existence of wormholes remains unsolved. A study on this topic has been done by Khatsymovsky in \cite{Khatsymovsky}. Also, the authors of \cite{Halilsoy,Ray}, investigated the possibility of the existence of wormholes in galactic halo regions. Recently, Bambi and  Stojkovic reviewed past and current efforts to search for astrophysical wormholes in the Universe in \cite{Bambi}.\\
In 1916, Flamm first realized the concept of wormhole \cite{Flamm} and acknowledged that the Schwarzschild black hole could provide a path for interstellar travels. Later, in 1935, Einstein and Rosen adopted this concept and constructed a hypothetical bridge or a wormhole mathematically \cite{Rosen}. But, the term `wormhole' was introduced for the first time in 1957 by Misner and Wheeler \cite{Misner}. Later it was admitted that the wormhole solutions do not construct a stable configuration - its `throat' shuts up too fast when subjected to even tiny perturbation \cite{Kruskal,Fuller,Eardley}. In 1988, Morris and Thorne first gave the idea of a traversable wormhole \cite{Thorne/1988}. They have provided some basic and desired properties of a wormhole for its traversability. In the context of GR, the primary component of the wormhole is the violation of energy conditions and the matter which disrespects the energy conditions (especially null energy conditions (NEC)) at or in the vicinity of the throat dubbed exotic matter \cite{Thorne/1988,Visser}. The exotic matter is used to define some cosmological observations, such as the behavior of the galactic rotation curves, the late-time accelerated expansion of the universe, and the mass discrepancy in clusters of galaxies. But, recently, some astrophysical observations have revealed that exotic matter is not required for an accelerated expansion of the Universe, and hence researchers considered phantom regions for wormholes study subsequent to this observation \cite{Lobo/2005,Sushkov,Parsaei,Moradpour11}. Further, such exotic matter can be required in both static \cite{Anabalon,Kuhfittig,Balakin} and dynamic \cite{Hansen,Diaz,Dehghani} wormhole cases.\\
In wormhole physics, the minimization of exotic matter is a big challenge. Therefore some techniques have been introduced in the literature, such as the ``cut and paste" method \cite{Visser1,Visser2} to minimize the usage of exotic matter, but this method is restricted to wormhole throat only. Also, Visser et al., \cite{Visser3} have developed a suitable measure called ``volume integral quantifier (VIQ)" to check the amount of exotic matter required for a traversable wormhole.\\
In the last few years, researchers have been growing curiosity about modified theories of gravity (MTG). Basically, MTGs are the geometrical generalizations of Einstein's GR in which cosmic acceleration can be gained by modifying the Einstein-Hilbert action integral. Many MTGs have been used to describe dark energy and, mainly, both early and late time acceleration expansion of the Universe. Since MTG helps us to explain cosmic expansion and further related concepts, it would be fascinating to experiment with the capacity of these theories to study astrophysical objects such as wormholes, compact stars, etc. Quadratic gravity \cite{Duplessis}, Born-Infeld theory \cite{Simeone,Aguirre,Shaikh}, curvature matter coupling \cite{Garcia,Garcia1}, Einstein-Cartan gravity \cite{Bronnikov1,Bronnikov2,Mehdizadeh1}, Rastall theory \cite{Moradpour}, and other modified gravities \cite{ad1,ad2,ad3,ad4,ad5,ad6,ad7} are few examples. Recently, in \cite{Tsukamoto,Shaikh3,Nedkova}, the authors have published some interesting research on the shadows of wormholes and Kerr-like wormholes. These types of works also discussed in $f(R)$ and $f(T)$ gravity as well (see Refs. \cite{Jamil1,Jamil2}). Further, one may study some interesting articles on wormhole geometries in different MTG such as in $f(R)$ gravity \cite{Fayyaz,Horvat11,Eiroa,Karakasis,Sossich}, $f(R,T)$ gravity \cite{Sahoo,Ahmad,Ilyas,Khurshudyan}, $f(T)$ gravity \cite{Harko/2012,Jamil/2013,Shamaila} and other modified theories gravity \cite{Ali,Nawazish,Godani,2,Tayde,Nazir,Nazir1}.\\
Over the years, significant growth has been witnessed in the extensions of GR \cite{Laurentis} involving torsion-based gravity \cite{Krssak}. Nevertheless, in 1999, the non-metricity theory came to light after the proposal of the so-called Symmetric Teleparallel Gravity \cite{Nester,Kalay,Conroy}. In this modified gravity, both curvature and torsion are set to zero; hence, gravitation is linked to the non-metricity tensor and affiliated to the nonmetricity scalar $Q$. Recently, Jimenez et al. has generalized this theory which has acquired significant attention from researchers, namely $f(Q)$ gravity \cite{Jimenez} where the gravitational field is expressed by the non-metricity scalar $Q$ only. This theory has successfully encountered various background and perturbation observational data such as the Supernovae type Ia (SNIa), Cosmic Microwave Background (CMB), Redshift Space Distortion (RSD), Baryonic Acoustic Oscillations (BAO), etc., \cite{Soudi,Banos,Salzano,Koivisto11}, and this conflict demonstrates that the $f(Q)$ gravity could challenge the $\Lambda$CDM model \cite{Anagnostopoulos}. Moreover, we could see the growing interest of $f(Q)$ gravity in the field of astrophysical objects as well. Black holes in $f(Q)$ gravity have been investigated in \cite{Fell}. In \cite{Zhai}, the authors have studied the application of the spherically symmetric configurations in f(Q) gravity. Further, the static and spherically symmetric solutions under anisotropic fluid for $f(Q)$ gravity have been discussed by Wang et al. in \cite{Wang2}. Hassan et al. \cite{Hassan1} have investigated wormhole geometries in $f(Q)$ gravity by choosing linear equations of state (EoS) and anisotropic relations. They have found exact solutions for the linear model and have confirmed a small amount of exotic matter required for a traversable wormhole via VIQ \cite{Hassan1}. Also, in \cite{Mustafa}, Mustafa et al. have obtained wormhole solutions from the Karmarkar condition and provide the possibility of obtaining traversable wormholes respecting the energy conditions. Recently, a class of static spherically symmetric solutions in $f(Q)$ gravity have been investigated in \cite{1}. For more applications of astrophysical objects in $f(Q)$ gravity, one may check the literature such as wormholes with charge \cite{Sokoliuk}, conformal symmetry \cite{Sahoo111}, and compact star \cite{Hassan3}.\\
It is well known that in Einstein's GR, for the wormhole to be traversable, we need to violate the Null Energy Condition (NEC), which confirms the presence of exotic matter at the wormhole throat. One practical example of such matters can be found in the Casimir effect. The Casimir effect appears if we put two parallel conducting plates in a vacuum. They attract themselves as the zero modes of the quantum field theory give rise to the energy between the plates. It was first discovered by \cite{casimir} and was later shown by \cite{lifshitz} in a different way. The experimental evidence of the Casimir effect is also known and has been shown in \cite{experiment,bressi}. In \cite{Garattini}, the author has recently presented a wormhole model probing the negative energy density because of the Casimir effect and explored the consequences of quantum weak energy conditions on the traversability of the wormhole.\\
The idea of the Generalized Uncertainty Principle (GUP) comes from the fact that in the quantum gravity theory, there is usually a fundamental length scale beyond which the resolution is not possible, such as in string theory, and the length of string, etc. It may be shown via renormalization group theory methods that such an elementary length scale is inhabitable, as demonstrated in \cite{rainbowsmolin}. There are many other phenomenological implications in quantum gravity theory if we allow a minimum length scale which has been discussed in detail in \cite{sabinegup}. \\
It is known that the uncertainty principle in quantum mechanics is given as follows, if $\hat{A}$ and $\hat{B}$ are two Hermitian operators, then the Uncertainty principle is defined as 
\begin{equation}
    \Delta A \Delta B \geq \frac{1}{2 i} \left\langle[A,B] \right \rangle,
\end{equation}
where $\Delta A=\sqrt{\left\langle \hat{A}^2 \right \rangle-\left(\left\langle \hat{A} \right \rangle \right)^2}.$\\
For position $x$ and momentum $p$ we can get the original position momentum uncertainty relation $\Delta x \Delta p\geq \frac{\hbar}{2}$,
where $\hbar$ is the plank constant.\\
Heisenberg's uncertainty principle has powerful experimental evidence. However, some serious problems occur when we try to incorporate it into GR, such as in classical GR, we get a singularity in the metric like in Schwarzschild solutions at $r=0$. Note that the uncertainty principle states that $\Delta x \Delta p\geq \frac{\hbar}{2}$. So by approximation, we can see the position momentum near the singularity would behave like $\Delta x \sim \frac{Const.}{\Delta p}$, but we know in natural units, the Schwarzschild solutions go as $\Delta x= 2G\Delta p$ as in natural unit $M$ (mass), $p$ (momentum), and $E$ (energy) have the same dimension. One way to make Schwarzschild singularity compatible with the Heisenberg uncertainty principle is to guess that the form of the momentum position relation has been something like $\Delta x \sim \frac{\hbar}{\Delta p}+G \Delta p$ 
or in natural units, we can roughly see the uncertainty principle becomes  $[x,p]=i\hbar(1+\lambda p^2)$
also, in the natural units $\Delta p_m\sim \sqrt{\frac{\hbar}{G}}$ and $\Delta x_m\sim \sqrt{\hbar G}$ so far as for the search for such a fundamental length scale goes. Various experiments have been proposed, like seeing the modified dispersion relation of a photon via gamma-ray burst \cite{ellis}; there is a phenomenological discussion in \cite{dasexp} on how the various experimental methods can probe the fundamental length scale of $\lambda$ like lamb shift $\lambda<10^{36}$, Landau level $\lambda<10^{56}$, tunneling $\lambda<10^{21}$. For various other experiments, one can see \cite{gupmain,sabinegup} and the reference therein. Corrections of the Casimir effect due to GUP are also well known and are discussed in \cite{gupcasimir}. The application of minimal time scale and GUP has already been successfully used in solving the Wheeler-Dewitt equation for the universe's evolution \cite{3}. In recent years GUP has also been used to find dispersion relation during Hawking radiation of Schwarzschild-de Sitter black holes \cite{4}, and can one can also get a limit of minimal length scale already knowing the blackhole evaporation formula using semiclassical quantum gravity.\\
The effect of GUP in the Casimir wormhole has been widely studied by Garattini in \cite{7}. Further, Samart et al. investigated the charged wormhole with and without GUP corrected Casimir wormhole in \cite{Channuie} for classical GR. These researches motivate us to study the effect of GUP Casimir wormholes in the recently proposed modified symmetric teleparallel gravity. Particularly, we consider two famous GUP relations such as the Kempf, Mangano and Mann (KMM) model and the Detournay, Gabriel and Spindel (DGS) model, and investigate a class of asymptotically flat wormhole solutions in the background of the effect of GUP corrected Casimir energy.\\
This article is organized as follows: We have introduced the basic formalism of $f(Q)$ gravity and constructed the Morris-Throne wormhole metric field equations for the affiliated gravity in Sec. \ref{sec2}. A brief review of the Casimir effect under the generalized uncertainty principle is presented in  Sec. \ref{sec3}. In Sec. \ref{sec4}, we construct the shape function by comparing the GUP corrected energy density with the wormhole metric's energy density and investigated the traversability conditions with different redshift functions under linear $f(Q)$ gravity. Further, a discussion on the energy conditions for both models under the linear $f(Q)$ form is placed in Sec. \ref{sec5}. Also, in \ref{sec5aa}, we used a numerical approach to study GUP corrected Casimir wormhole for the quadratic case, showed the possible form for the shape function, and the ADM mass is also examined in Sec. \ref{sec6}. Furthermore, to investigate the amount of exotic matter necessary for wormhole maintenance, we used the Volume Integral Quantifier, presented in \ref{sec7}. Finally, we conclude our results in the last section.

\section{Basic formulation of wormholes and field equations in $f(Q)$ gravity}\label{sec2}
Here we consider the static and spherically symmetric Morris-Thorne wormhole metric  \cite{Visser,Thorne/1988}, defined by
\begin{equation}\label{3a}
ds^2=U(r)dt^2-V(r)dr^2-r^2d\Omega^2,
\end{equation}
where, $d\Omega^2=d\theta^2+\text{sin}^2\theta d\Phi^2$, $U(r)=e^{\phi(r)}$ and $V(r)=\left(1-\frac{b(r)}{r}\right)^{-1}$.
The function $b(r)$ designated as the shape function is used to define the shape of the wormholes. The expression $\phi(r)$ is the redshift function, and it is related to the gravitational redshift. To avoid the event horizon, $\phi(r)$ must be finite everywhere. Besides this, to have wormhole geometry, $b(r)$ should satisfy the flaring-out condition, which is given by $(b-b'r)/b^2>0$ \cite{Thorne/1988} and at the throat $b(r_0)=r_0$ ($r_0$ is the throat radius), the condition $b^{\,\prime}(r_0)<1$ is imposed. Further, the asymptotic flatness condition, that is, the limit $\frac{b(r)}{r}\rightarrow 0$ as $r\rightarrow \infty$ is also required. In GR, satisfying the above conditions may confirm the presence of exotic matter at the throat of the wormhole.\\
Now, we are going to briefly present some generalities about the $f(Q)$ gravity. The action for this gravity is given by \cite{Jimenez}
\begin{equation}\label{6a}
\mathcal{S}=\int\frac{1}{2}\,f(Q)\sqrt{-g}\,d^4x+\int \mathcal{L}_m\,\sqrt{-g}\,d^4x\, ,
\end{equation}
where, $\mathcal{L}_m$ denoted as Lagrangian density of matter and $g=\text{Det}[g_{\mu\nu}]$. Here $f(Q)$ is the arbitrary function of the non-metricity scalar $Q$.\\
The non-metricity tensor may be denoted by\\
\begin{equation}\label{6b}
Q_{\lambda\mu\nu}=\bigtriangledown_{\lambda} g_{\mu\nu}=\partial_\lambda g_{\mu\nu}-\Gamma^\beta_{\,\,\,\lambda \mu}g_{\beta \nu}-\Gamma^\beta_{\,\,\,\lambda \nu}g_{\mu \beta},
\end{equation}
where, $\Gamma^\beta_{\,\,\,\mu\nu}$ is the metric affine connection.\\
Also its independent traces may be reads as
\begin{equation}
\label{6d}
Q_{\alpha}=Q_{\alpha}\;^{\mu}\;_{\mu},\; \tilde{Q}_\alpha=Q^\mu\;_{\alpha\mu}.
\end{equation}
The non-metricity scalar is represented as \cite{Jimenez}
\begin{equation}
\label{6e}
Q=-Q_{\alpha\mu\nu}\,P^{\alpha\mu\nu},
\end{equation}
where, $P^\alpha\;_{\mu\nu}$ is the non-metricity conjugate and it may be defined by
\begin{multline}\label{6c}
P^\alpha\;_{\mu\nu}=\frac{1}{4}\left[-Q^\alpha\;_{\mu\nu}+2Q_{(\mu}\;^\alpha\;_{\nu)}+Q^\alpha g_{\mu\nu}-\tilde{Q}^\alpha g_{\mu\nu}\right.\\\left.
-\delta^\alpha_{(\mu}Q_{\nu)}\right].
\end{multline}
Also, the standard energy-momentum tensor can be written as
\begin{equation}\label{6g}
T_{\mu\nu}=-\frac{2}{\sqrt{-g}}\frac{\delta\left(\sqrt{-g}\,\mathcal{L}_m\right)}{\delta g^{\mu\nu}}.
\end{equation}
Now, to obtain the field equations for this theory, we vary the action \eqref{6a} with regard to the metric tensor $g_{\mu\nu}$, the motion equations
\begin{multline}\label{6f}
\frac{2}{\sqrt{-g}}\bigtriangledown_\gamma\left(\sqrt{-g}\,f_Q\,P^\gamma\;_{\mu\nu}\right)+\frac{1}{2}g_{\mu\nu}f \\
+f_Q\left(P_{\mu\gamma i}\,Q_\nu\;^{\gamma i}-2\,Q_{\gamma i \mu}\,P^{\gamma i}\;_\nu\right)=-T_{\mu\nu},
\end{multline}
where $f_Q=\frac{df}{dQ}$.
Also, by varying the action over the connection, one obtains
\begin{equation}\label{6h}
\bigtriangledown_\mu \bigtriangledown_\nu \left(\sqrt{-g}\,f_Q\,P^\gamma\;_{\mu\nu}\right)=0.
\end{equation}
In this study, we consider the diagonal energy-momentum tensor for an anisotropic fluid of the form
\begin{equation}\label{3b}
T^{\mu}_{\nu}=\text{diag}[\rho,\,-P_r,\,-P_t,\,-P_t],
\end{equation}
where, $\rho$ is the energy density. $P_r$ and $P_t$ denote the radial and tangential pressure, respectively.\\
For the line element \eqref{3a}, we are able to find the following non-metricity scalar from Eq. \eqref{6e}
\begin{equation}\label{6i}
Q=-\frac{b}{r^2}\left[\frac{r b'-b}{r (r-b)}+\phi^{'}\right].
\end{equation}
Now, the field equations for the wormhole metric \eqref{3a} with anisotropic matter source \eqref{3b} in modified symmetric teleparallel gravity may be obtained as
\begin{multline}
\label{11}
\rho =\frac{(r-b)}{2 r^3}\left[2\,r\,f_{QQ}Q^{'}\frac{b}{r-b}\right.\\\left.
+f_Q \left(\frac{(2 r-b) \left(r b^{'}-b\right)}{(r-b)^2}+\frac{b \left(r \phi^{'}+2\right)}{r-b}\right)+\frac{f r^3}{r-b}\right],
\end{multline}
\begin{multline}
\label{12}
P_r=-\frac{\left(r-b\right)}{2 r^3}\left[2\,r\,f_{QQ}Q^{'}\frac{b}{r-b}\right.\\\left.
+f_Q \left(\frac{b}{r-b}\left(\frac{r b{'}-b}{r-b}+r \phi^{'}+2\right)-2 r \phi^{'}\right)+\frac{f r^3}{r-b}\right],
\end{multline}
\begin{multline}
\label{13}
P_t=-\frac{\left(r-b\right)}{4 r^2}\left[-2\,r\,\phi^{'} f_{QQ} Q^{'}\right.\\\left.
+ f_Q \left(\frac{\left(r b^{'}-b\right)}{r (r-b)}\left(\frac{2 r}{r-b}+r \phi^{'}\right)+\frac{2 (2 b-r) \phi^{'}}{r-b}\right.\right.\\\left.\left.
-r \left(\phi^{'}\right)^2-2 r \phi^{''}\right)+\frac{2 f r^2}{r-b}\right],
\end{multline}
where ${'}$ represents $\frac{d}{dr}$.\\
\section{The Casimir effect under the GUP}
\label{sec3}
\subsection{Casimir effect}
One of the natural sources of exotic matter which naturally comes for quantization of field is Casimir energy. It is well known that if we keep two parallel conducting plates in close proximity, they get attracted, and the energy is given by the formula below,
\begin{equation}\label{4a1}
   E(a)=-\frac{\pi^2}{720}\frac{S}{a^3}.
 \end{equation}
This formula was first derived in \cite{casimir} and independently by \cite{lifshitz}, and later experimentally verified in \cite{experiment,bressi}.\\
One can show that the expression of Eq. \eqref{4a1} comes from summing over the normal modes of the field and adequately regularizing the sum. One can do the regularization in two ways: first, by introducing a cutoff limit \cite{zee}, and second via an analytic continuation of the Riemann Zeta function \cite{paddy}, both of which lead to the same answer.\\
It is known that the stability of a wormhole requires the NEC violation, and the exotic matter is necessary for the stability of the wormhole. Casimir energy can be used for such exotic matter sources as it has been studied in detail in various ``Casimir wormhole" articles (See Refs. \cite{Bezerra,Fuenmayor,Khabibullin}). However, on such a small quantum scale, it is not just necessary for the vacuum fluctuation but also for the fundamental length scale that gets important. So we need to use the fact that there is a natural length scale associated with the fundamental theory underlying quantum gravity; as we will see in the following subsection that such a fundamental length scale will give rise to the generalized uncertainty principle. In fact, the order of magnitude of such correction in the Casimir magnitude effect is quite comparable with the general Casimir force. 
\subsection{Generalized Uncertainty Principle}
As stated earlier, the existence of a minimum length scale leads to modifying the uncertainty principle. Also, in the definition of position and momentum, various issues of GUP are discussed in \cite{kmm,gupmath,gupmath2,dgs}.\\
Here we are only interested in the effect of Casimir energy due to GUP. We follow \cite{gupcasimir} for the correction of Casimir energy in GUP from this paper. We also note that position and momentum are no longer conjugate variables in the classical sense, so we can not talk about the position eigenspace as actual physical position since we have changed the position momentum relation. One way to still talk about the position as the states projected onto the maximally localized state is also known as the ``quasi position representation" as discussed in \cite{kmm}.
However, there are mainly two ways to get the physical position states or ``maximally localized states", one is the canonical approach by KMM \cite{kmm}, and later DGS \cite{dgs} found that all squeezed states which represent the ``maximally localized states" can not be found by this method, so they have used the variational program to find that.\\
We have used both methods of GUP to see the effect of change in the Casimir force and how it affects our wormhole solutions. We note that from \cite{gupcasimir} that in the n-special dimension, GUP can be defined as 
\begin{equation}
[x_i,p_j]=i\hbar[f(p^2)\delta_{ij}+g(p^2)p_ip_j].
\end{equation}
We also note that these are the only options in the first order due to spherical symmetry. Also, we note that the form of $f$ and $g$ are not arbitrary, as discussed in \cite{kmm}.\\
We can see the quantum state, which may be written as 
\begin{equation}\label{c1}
    \psi_r=\frac{1}{(\sqrt{2\pi\hbar})^3}\Omega(p)\exp\left({-\frac{i}{\hbar}[k(p).r-\hbar\omega(p) t]}\right).
\end{equation}
where the functional form of $\omega(p)$ denotes the dispersion relation that can be found in theory. Here $\Omega$ and $k$ are the measure and the wave vector, respectively.\\
Below we discuss two of the most popular ways of GUP that is KMM \cite{kmm} which uses squeezed state and DGS \cite{dgs} which uses variational principle and is a little more useful in the general case.\\
The various field theoretic formulations of GUP, as well as the issue of ultraviolet divergence issues, etc., can be found in \cite{regula,gupcasimir}.
\subsubsection{KMM Model}\label{subsec1}
The specific form of these states depends on the number of dimensions and the specific model considered. There are at least two different approaches to constructing maximally localized states in the literature: the procedure proposed by Kempf, Mangano, and Mann (KMM) \cite{kmm}. This model corresponds to the choice of the generic functions $f(\hat{p}^2)$ and $g(\hat{p}^2)$ given in \cite{regula}.
\begin{equation}\label{c2}
 f(\hat{p}^2)=\frac{\lambda \hat{p}^2}{\sqrt{1+2\lambda \hat{p}^2}-1}, \quad g(\hat{p}^2)=\lambda.
\end{equation}
From now onwards, we shall remove the hat over the operator. The KMM construction of maximally localized states gives Eq. \eqref{c1} the following functions:
\begin{equation}
    \kappa_i(p)=\left(\frac{\sqrt{1+2\lambda p^2}-1}{\lambda p^2}\right)p_i,
\end{equation}
\begin{equation}
    \omega(p)=\frac{pc}{\hbar}\left(\frac{\sqrt{1+2\lambda p^2}-1}{\lambda p^2}\right),
\end{equation}
\begin{equation}
    \Omega(p)=\left(\frac{\sqrt{1+2\lambda p^2}-1}{\lambda p^2}\right)^{\frac{\delta}{2}},
\end{equation}
where $n$ denote the number of space-time dimensions, and $\delta=1+\sqrt{1+\frac{n}{2}}$ represents the KMM approach. Now we could determine the identity operator from the scalar product of maximally localized states
\begin{equation}
    \int \frac{d^n p}{\sqrt{1+2\lambda p^2}}\left(\frac{\sqrt{1+2\lambda p^2}-1}{\lambda p^2}\right)^{(n+\delta)}\vert p\rangle\langle p \vert=1.
 \end{equation}

\subsubsection{DGS Model}
As defined earlier, various maximally localized states may correspond to a given choice of generic functions \eqref{c2}. The DGS \cite{dgs} maximally localized forms are provided by Eq. \eqref{c1} with
\begin{equation}
    \kappa_i(p)=\left(\frac{\sqrt{1+2\lambda p^2}-1}{\lambda p^2}\right)p_i,
\end{equation}
\begin{equation}
    \omega(p)=\frac{pc}{\hbar}\left(\frac{\sqrt{1+2\lambda p^2}-1}{\lambda p^2}\right),
\end{equation}
\begin{multline}
\Omega(p)=\left[\Gamma\left(\frac{3}{2}\right)\left(\frac{2\sqrt{2}}{\pi\sqrt{\lambda}}\right)^{\frac{1}{2}}\right]\left[\frac{1}{p}\frac{\lambda p^2}{\sqrt{1+2\lambda p^2}-1}\right]^{\frac{1}{2}}\textbf{J}_{\frac{1}{2}}\\
\times \left[\frac{\pi \sqrt{\lambda}}{\sqrt{2}}\left(\frac{\sqrt{1+2\lambda p^2}-1}{\lambda p^2}\right)p\right],
\end{multline}
where $\textbf{J}_{\frac{1}{2}}$ is the Bessel function of the first kind. Now on solving the above expression, one can obtain
\begin{multline}
\Omega(p)=\frac{\sqrt{2}}{\pi}\frac{\sqrt{\lambda}p}{\sqrt{1+2\lambda p^2}-1}\\
\times \text{sin}\left[\frac{\sqrt{2}\pi(\sqrt{1+2\lambda p^2}-1)}{2\sqrt{\lambda}p}\right].
\end{multline}
The modified identity operator for the momentum eingestates $\vert p\rangle$ for this case
\begin{equation}
    \int \frac{d^n p}{\sqrt{1+2\lambda p^2}}\left(\frac{\sqrt{1+2\lambda p^2}-1}{\lambda p^2}\right)^{n}\vert p\rangle\langle p \vert=1.
 \end{equation}

\subsection{GUP corrected energy density}
In \cite{gupcasimir}, the authors have employed the concept of minimal length and GUP to obtain the finite energy between the plane plates. They have derived the Hamiltonian and the corrections to the Casimir energy due to the minimal length. Up to a first order correction term in the minimal uncertainty parameter $\lambda$, the Casimir energy for the two different cases of construction of maximally localized states are obtained as
\begin{equation}\label{51}
E=-\frac{\pi^2 S}{720}\frac{\hbar}{a^3}\left[1+\Lambda_i\left(\frac{\hbar \sqrt{\lambda}}{a}\right)^2\right],
\end{equation}
where $S$ is the surface area of the plates and $a$ is the separation between them. $\Lambda$ is a constant where $i=1,\,2$. In particular we have the following two cases:
$$\Lambda_1=\pi^2\left(\frac{28+3\sqrt{10}}{14}\right)\quad (\text{for KMM model}),$$
$$\Lambda_2=\left(\frac{4\pi^2(3+\pi^2)}{21}\right) \quad (\text{for DGS model}).$$
Then the force can be obtained with the computation of
\begin{equation}\label{52}
F=-\frac{dE}{da}=-\frac{3\pi^2 S}{720}\frac{\hbar}{a^4}\left[1+\frac{5}{3}\Lambda_i\left(\frac{\hbar \sqrt{\lambda}}{a}\right)^2\right].
\end{equation}
Thus, we get the formula for pressure
\begin{equation}\label{53}
P=\frac{F}{S}=-\frac{3\pi^2}{720}\frac{\hbar}{a^4}\left[1+\frac{5}{3}\Lambda_i\left(\frac{\hbar \sqrt{\lambda}}{a}\right)^2\right]=\omega\rho.
\end{equation}
From the above equation, we can see that EoS can be defined by putting $\omega=3$. Now we can see that in natural units, the GUP-corrected energy density becomes
\begin{equation}\label{54}
\rho=-\frac{\pi^2}{720}\frac{1}{a^4}\left[1+\frac{5}{3}\Lambda_i\left(\frac{\lambda}{a^2}\right)\right].
\end{equation}
Setting $\lambda=0$, we obtain the usual Casimir result.




\section{GUP corrected Casimir wormholes for the linear $f(Q)=\alpha\,Q+\beta$ model}\label{sec4}
\begin{figure*}[h]
    \centering
    \includegraphics[scale=0.6]{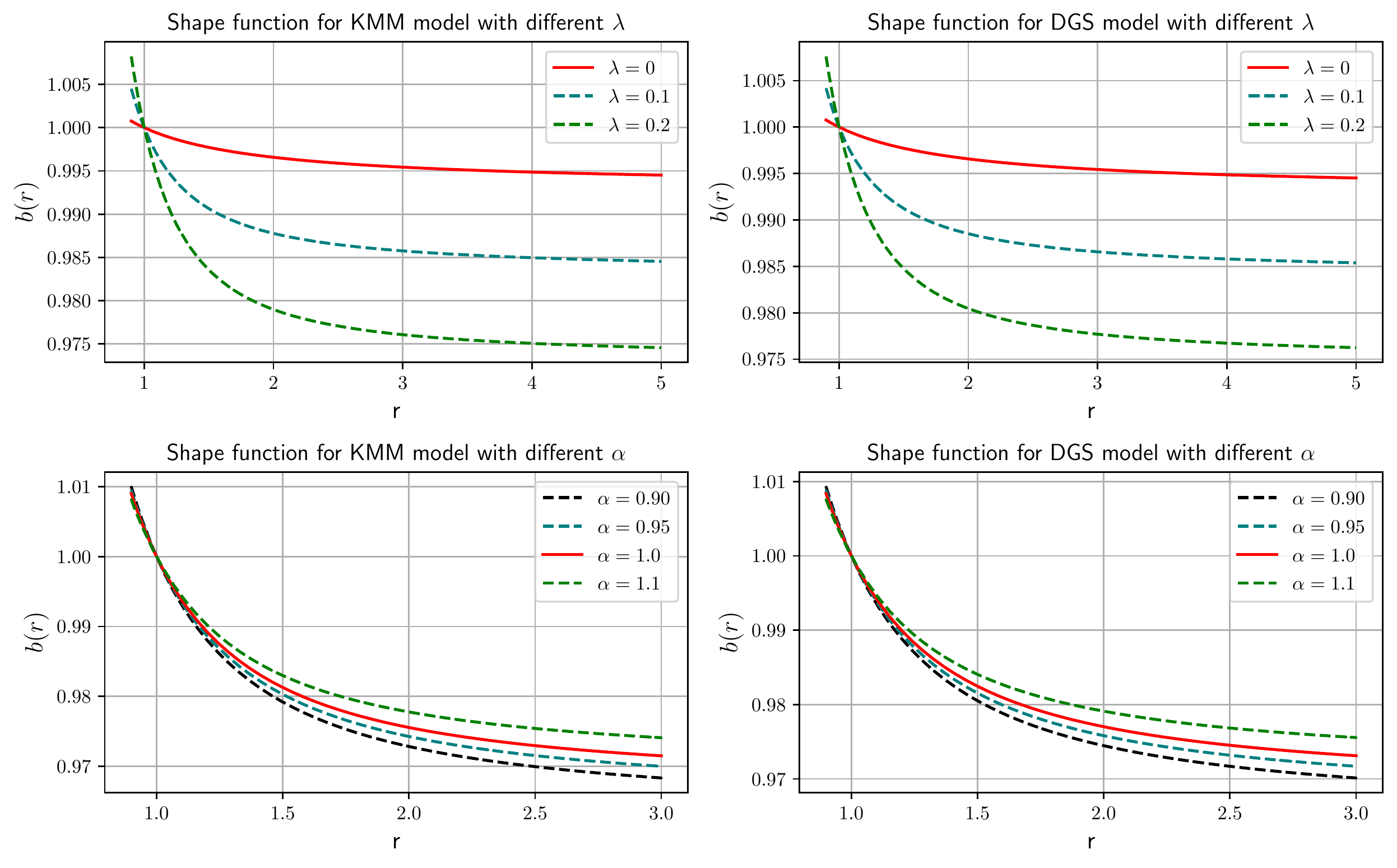}
    \caption{Shape functions for KMM and DGS models against radial distance $r$ with $r_0=1$.  We fix $\alpha=2$ for the upper half and $\lambda=0.1$ for the lower half in the figures. Also, note that on the upper half figures $\lambda=0$ corresponds to the usual Casimir wormhole, and on the lower half figures $\alpha=1$ corresponds to the GR case.}
    \label{fig:1}
\end{figure*}

\begin{figure*}[h]
    \centering
    \includegraphics[scale=0.6]{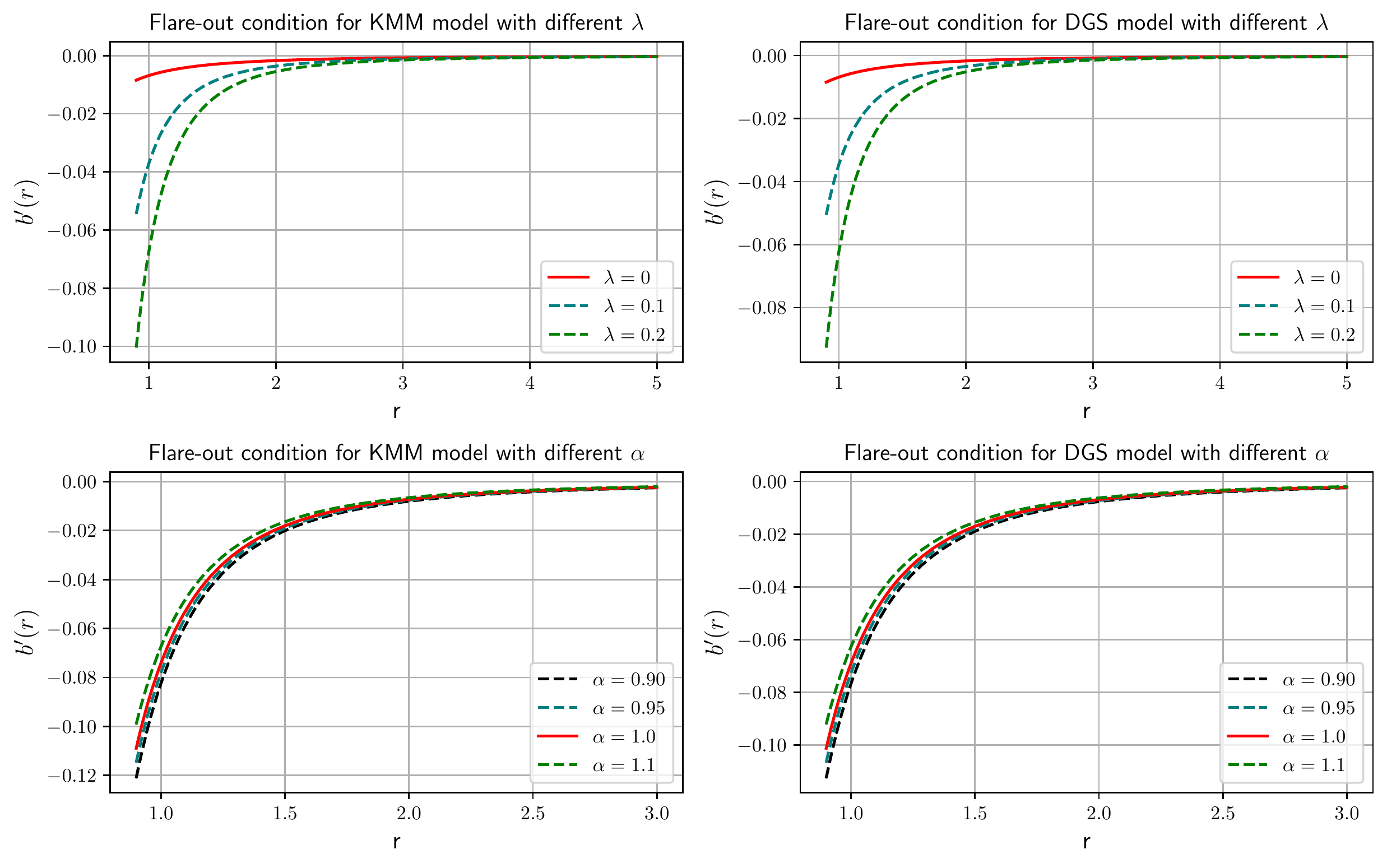}
    \caption{Flare-out conditions for KMM and DGS models against radial distance $r$ with $r_0=1$.  We fix $\alpha=2$ for the upper half and $\lambda=0.1$ for the lower half in the figures. Also, note that on the upper half figures $\lambda=0$ corresponds to the usual Casimir wormhole, and on the lower half figures $\alpha=1$ corresponds to the GR case.}
    \label{fig:2}
\end{figure*}

In this segment, we assume the simplest linear functional form of $f(Q)$ gravity, such as $f(Q)=\alpha Q+\beta$, where $\alpha$ (\textbf{the bound has been motivated from \cite{Solanki}}) and $\beta$ are model parameters. Note that the above model can be reduced to GR if we consider $\alpha=1$ and $\beta=0$. This particular form is derived from the most general power law form $f(Q)=\alpha Q^{n+1}+\beta$ \cite{Parbati}. With this specific linear model, Solanki et al., \cite{Solanki} have investigated the late-time cosmic acceleration without invoking any dark energy component in the matter part. For this linear model, our general field equations (\ref{11}-\ref{13}) reduces to\\
\begin{equation}
\label{14}
\rho=\frac{\alpha  b'}{r^2}+\frac{\beta }{2},
\end{equation}
\begin{equation}
\label{15}
P_r=\frac{1}{r^3}\left[2 \alpha  r \left(r-b\right) \phi '-\alpha  b\right]-\frac{\beta }{2},
\end{equation}
\begin{multline}
\label{16}
P_t=\frac{1}{2 r^3}\left[\alpha  \left(r \phi '+1\right) \left(-r b'+2 r (r-b) \phi '+b\right)\right]\\
+\frac{\alpha  (r-b) \phi ''}{r}-\frac{\beta }{2}.
\end{multline}
Further, to obtain the shape function of the GUP corrected energy density, we replace the plate separation distance $a$ by the radial distance $r$ in Eq. \eqref{54}. In that case, we can rewrite the energy density from Eq. \eqref{54} as
\begin{equation}\label{54aa}
\rho=-\frac{\pi^2}{720}\frac{1}{r^4}\left[1+\frac{5}{3}\Lambda_i\left(\frac{\lambda}{r^2}\right)\right].
\end{equation}
Now comparing Eqs. \eqref{14} and \eqref{54aa}, and solving the differential equation for shape function $b(r)$, we obtain
\begin{equation}\label{5111}
b(r)=-\frac{1}{2160 \alpha }\left[-\frac{5 \pi ^2 \lambda\Lambda_i}{3 r^3}+360 \beta  r^3-\frac{3 \pi ^2}{r}\right]+c_1,
\end{equation}
where $c_1$ is the integrating constant, and to calculate it, we apply throat condition $b(r_0)=r_0$ in the above equation, we get
\begin{equation}\label{5222}
c_1=r_0+\frac{1}{2160 \alpha }\left[-\frac{5 \pi ^2 \lambda\Lambda_i}{3 r_0^3}+360 \beta  r_0^3-\frac{3 \pi ^2}{r_0}\right].
\end{equation}
Inserting Eq. \eqref{5222} into Eq. \eqref{5111}, we obtain the final version of shape function $b(r)$ as follows
\begin{multline}\label{61}
b(r)=r_0+\frac{\xi_1}{5}\left(\frac{1}{r}-\frac{1}{r_0}\right)+\frac{\xi_1\lambda\Lambda_i}{9}\left(\frac{1}{r^3}-\frac{1}{r_0^3}\right)\\
+\frac{\beta}{6\alpha}\left(r_0^3-r^3\right),
\end{multline}
where 
\begin{equation}\label{62}
\xi_1=\frac{\pi^2}{144\,\alpha}.
\end{equation}
It may be observed that the above equation is not asymptotically flat, that is, for $r\rightarrow \infty$, $\frac{b(r)}{r}\nrightarrow 0$. It happens because of the fourth term of the above equation. For $\beta \rightarrow 0$, it will satisfy the flatness condition. From now on, we consider $\beta=0$ in this work. The last equation reduces to
\begin{equation}\label{63}
b(r)=r_0+\frac{\xi_1}{5}\left(\frac{1}{r}-\frac{1}{r_0}\right)+\frac{\xi_1\lambda\Lambda_i}{9}\left(\frac{1}{r^3}-\frac{1}{r_0^3}\right).
\end{equation}
The GUP correction term is proportional to the minimal uncertainty parameter $\lambda$. Clearly, in the limit $\lambda \rightarrow 0$, the shape function reduces to that of the Casimir wormhole \cite{Kazuharu}.\\
In Figs. \ref{fig:1} and \ref{fig:2}, we have depicted the behavior of shape functions for both models. It can be observed from Fig. \ref{fig:1} that for the increasing values of $\lambda$, the shape function $b(r)$ shows positively decreasing behavior, whereas, for the increase of $\alpha$, it shows increasing behavior. Also, from Fig. \ref{fig:2}, it is confirmed that the flaring-out condition is satisfied in the vicinity of the wormhole throat under asymptotic background.\\
For the GUP-corrected Casimir wormhole, the field equations (\ref{14}-\ref{16}) can be read as
\begin{equation}\label{7777}
\rho(r)=-\frac{\pi ^2 \left(5 \Lambda_i \lambda +3 r^2\right)}{\mathcal{D}_1},
\end{equation}
\begin{multline}\label{8888}
P_r(r)=\frac{1}{3 \mathcal{D}_1 r_0^3}\left[r (r-r_0) \phi ' \mathcal{D}_2+\pi ^2 \left(5 \Lambda_i \lambda  \left(r^3-r_0^3\right)\right.\right.\\\left.\left.
+9 r^2 r_0^2 (r-r_0)\right)-\mathcal{D}_3\right],
\end{multline}
\begin{multline}\label{9999}
P_t(r)=\frac{1}{12 \mathcal{D}_1 r_0^3}\left[r \left(\phi ' \left(r (r-r_0) \phi '\mathcal{D}_2+5 \pi ^2 \Lambda_i \lambda \right.\right.\right.\\\left.\left.\left.
\times \left(r^3+2 r_0^3\right)+\mathcal{D}_4\right)+2 r (r-r_0) \phi '' \mathcal{D}_2\right)+2 \pi ^2 \right.\\\left.
\times \left(-5 \Lambda_i \lambda  \left(r^3-4 r_0^3\right)-9 r^2 r_0^2 (r-2 r_0)\right)+2\mathcal{D}_3\right],
\end{multline}
where,
\begin{equation}
\mathcal{D}_1=2160 r^6,
\end{equation}
\begin{equation}\label{65aa}
\mathcal{D}_2=5 \pi ^2 \Lambda_i \lambda  \left(r^2+r r_0+r_0^2\right)+9 r^2 r_0^2 \left(720 \alpha  r r_0+\pi ^2\right),
\end{equation}
\begin{equation}\label{65a}
\mathcal{D}_3=6480 \alpha  r^3 r_0^4,
\end{equation}
\begin{equation}\label{65aaa}
\mathcal{D}_4=9 r^3 r_0^2 \left(720 \alpha  r_0 (2 r-r_0)+\pi ^2\right).
\end{equation}
Now, we shall present our study in the following subsections with the above components of generalized field equations.

\subsection{Case-I: $\phi(r)=k$}
In this subsection, we consider $\phi(r)=k$, where $k$ is any constant and hence $\phi^{'}(r)=0$.\\
For this case, the wormhole metric can be read as
\begin{multline}
ds^2=-e^{k}dt^2+\frac{dr^2}{1-\frac{r_0}{r}+\frac{\xi_1}{5r}\left(\frac{1}{r}-\frac{1}{r_0}\right)+\frac{\xi_1\lambda\Lambda_i}{9r}\left(\frac{1}{r^3}-\frac{1}{r_0^3}\right)}\\
+r^2\,d\theta^2+r^2\text{sin}^2\theta\,d\Phi^2,
\end{multline}
where, $\xi_1$ is defined in Eq. \eqref{62}.\\
Now we derive the equation of state (EoS) for the radial pressure defined by
\begin{equation}\label{64}
    P_r(r)=\omega_r(r)\rho(r),
\end{equation}
where $\omega$ is the EoS parameter which is a function of $r$.\\
Considering Eqs. (\ref{14}-\ref{16}) with shape function \eqref{63} under constant redshift function (zero tidal force), we obtain
\begin{equation}
\omega_r=\frac{\pi ^2 \left(5 \Lambda_i \lambda  \left(r_0^3-r^3\right)+9 r^2 r_0^2 (r_0-r)\right)+\mathcal{D}_3}{\mathcal{K}_1},
\end{equation}
where
\begin{equation}\label{65}
\mathcal{K}_1=3 \pi ^2 r_0^3 \left(5 \Lambda_i \lambda +3 r^2\right).
\end{equation}
\begin{figure}[H]
    \includegraphics[scale=0.57]{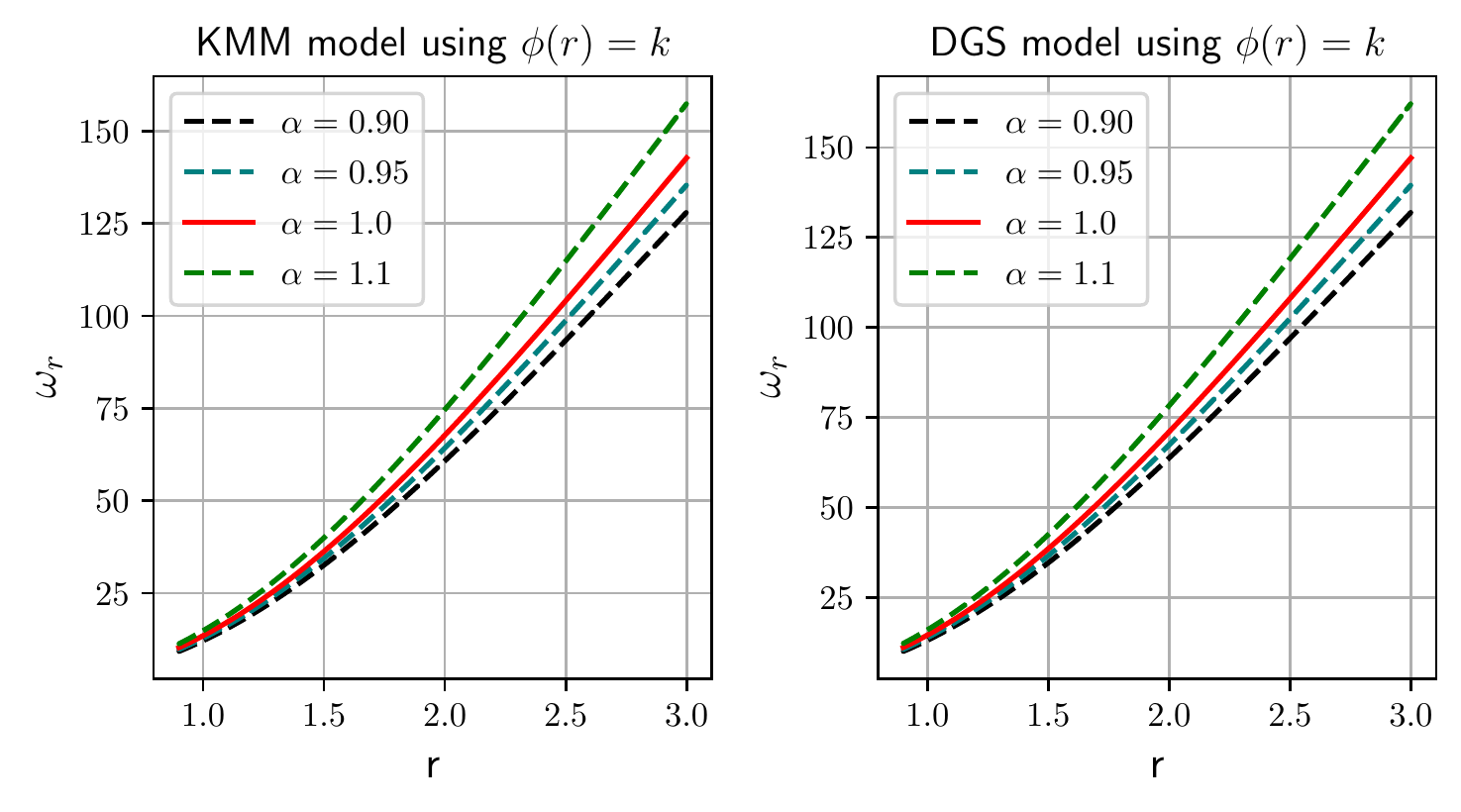}
    \caption{EoS parameter $\omega_r$ for KMM (left) and DGS (right) model using $\phi(r)=k$ for different $\alpha$ with $r_0=1$ and $\lambda=0.1$. In the figure, $\alpha=1$ corresponds to the GR case.}
    \label{fig:3}
\end{figure}
The behavior of radial EoS parameter $\omega_r$ for both KMM and DGS models has been illustrated in Fig. \ref{fig:3}. It is observed that radial EoS parameter $\omega_r$ increases positively with the increased values of $\alpha$ and radial distance $r$.

\subsection{Case-II: $\phi(r)=\frac{k}{r}$}
\begin{figure}[h]
    \includegraphics[scale=0.57]{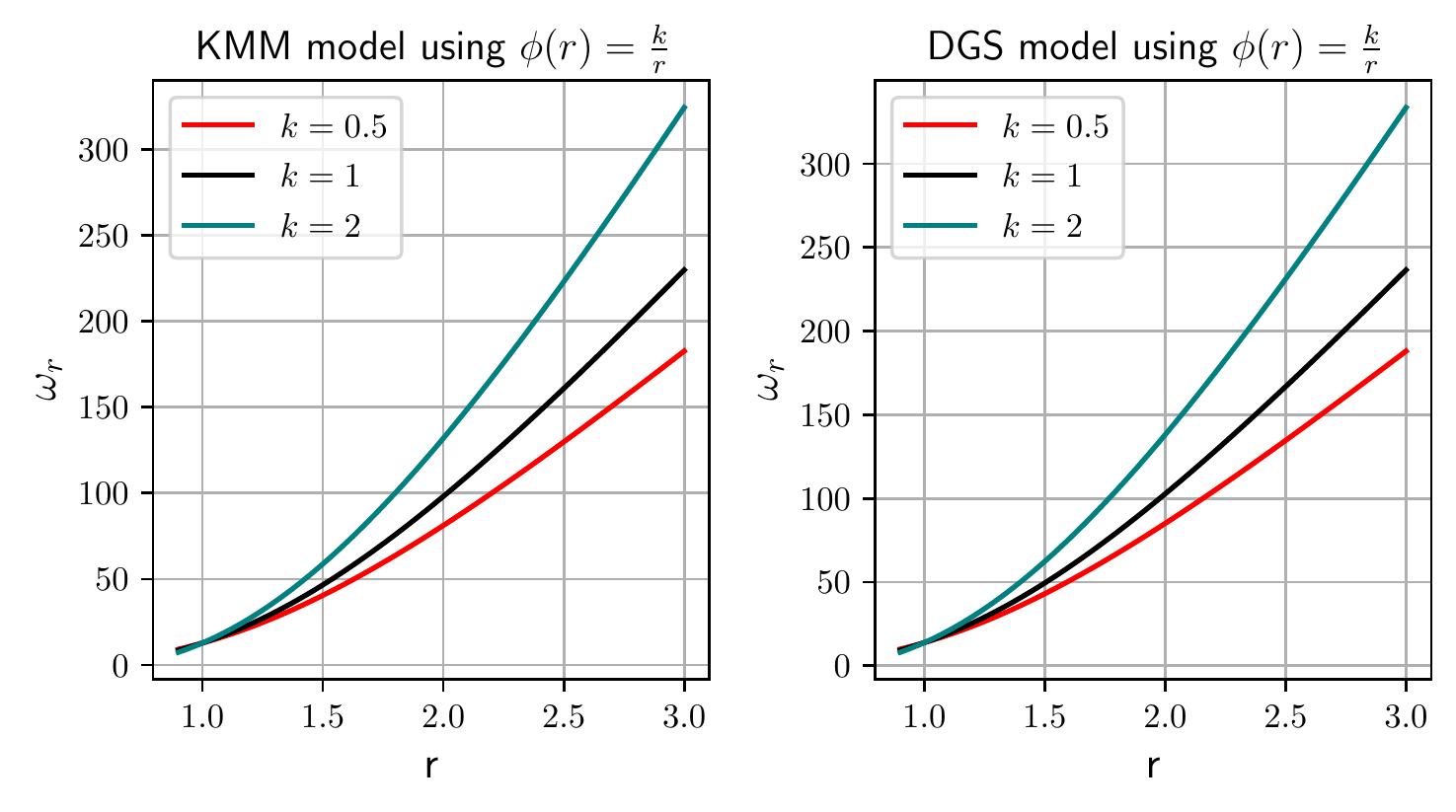}
    \caption{EoS parameter $\omega_r$ for KMM (left) and DGS (right) model using $\phi(r)=\frac{k}{r}$ for three different values of $k$ with $r_0=1$, $\lambda=0.1$ and $\alpha=0.95$.}
    \label{fig:4}
\end{figure}

We shall start our investigation by considering the radial EoS relation \eqref{64}. From the field equations \eqref{14} and \eqref{15}, we can determine the redshift function
\begin{equation}\label{66}
\phi^{'}(r)=\frac{\omega_rb^{'}r+b}{r(r-b)},
\end{equation}
for the redshift function $\phi(r)=\frac{k}{r}$, we are able to find the EoS parameter
\begin{equation}
\omega_r(r)= -\frac{k(r-b)+br}{b^{'}r^2}.
\end{equation}
with shape function \eqref{63}
\begin{multline}
\omega_r=\frac{1}{r\mathcal{K}_1}\left[5 \pi ^2 \Lambda_i \lambda  (k-r) \left(r^3-r_0^3\right)+9 r^2 r_0^2\right.\\\left.
\times \left(720 \alpha  r r_0 (k (r-r_0)+r r_0)+\pi ^2 (k-r) (r-r_0)\right)\right].
\end{multline}
at wormhole throat, the last equation reduces to
\begin{equation}
\omega_r\mid_{r=r_0}=\frac{2160 \alpha  r_0^4}{\pi ^2 \left(5 \Lambda_i \lambda +3 r_0^2\right)}.
\end{equation}
The behavior of the EoS parameter for both models is shown in Fig. \ref{fig:4}.

\subsection{Case-III: $\phi(r)=\ln \left(\frac{\sqrt{\eta^2+r^2}}{r}\right)$}
\begin{figure}[h]
    \includegraphics[scale=0.57]{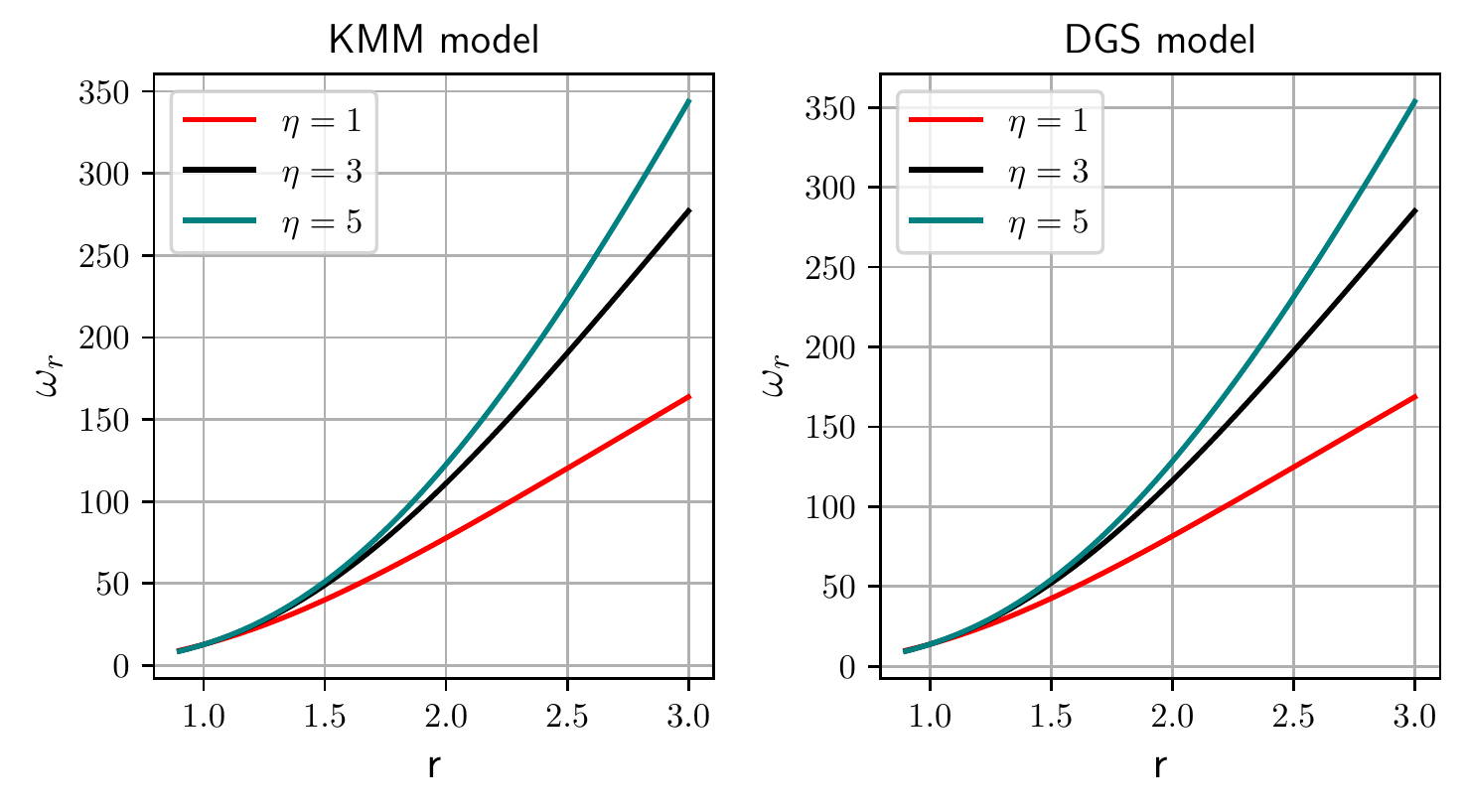}
    \caption{EoS parameter $\omega_r$ for KMM (left) and DGS (right) model using $\phi(r)=\frac{1}{2}\text{log}\left(1+\frac{\eta^2}{r^2}\right)$ for different $\eta$ with $r_0=1$, $\lambda=0.1$ and $\alpha=0.95$.}
    \label{fig:5a}
\end{figure}

\begin{figure}[h]
    \includegraphics[scale=0.57]{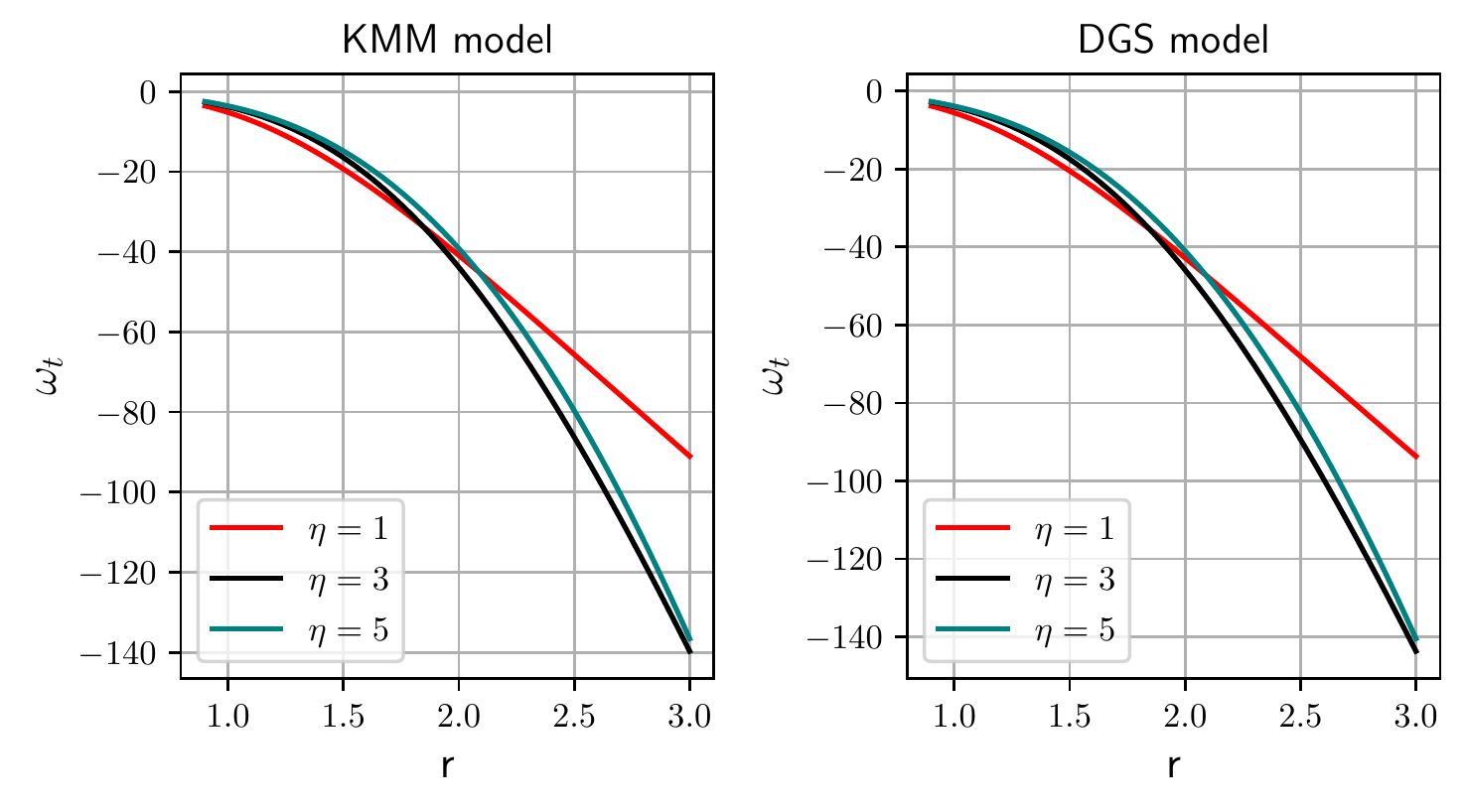}
    \caption{EoS parameter $\omega_t$ for KMM (left) and DGS (right) model using $\phi(r)=\frac{1}{2}\text{log}\left(1+\frac{\eta^2}{r^2}\right)$ for different $\eta$ with $r_0=1$, $\lambda=0.1$ and $\alpha=0.95$.}
    \label{fig:8z}
\end{figure}

For this specific redshift function, our wormhole metric can be read as
\begin{multline}
 ds^2=-\left(1+\frac{\eta^2}{r^2}\right)^{\frac{1}{2}}dt^2+V(r)dr^2+r^2\,d\Omega^2,
\end{multline}
where $\eta$ is any positive parameter. For this non-constant redshift function, from Eq. \eqref{64},  we could find the EoS parameter $\omega_r$ as
\begin{equation}
\omega_r(r)=-\frac{r b+\eta ^2}{\left(\eta ^2+r^2\right) b'},
\end{equation}
and for the shape function \eqref{63}, the above equation reduces to
\begin{multline}
\omega_r(r)=\frac{1}{\mathcal{K}_2}\left[5 \pi ^2 \Lambda_i \lambda  r^2 \left(r_0^3-r^3\right)+9 r^4 r_0^2 \right.\\\left.
\times \left(720 \alpha  r_0 \left(\eta ^2+r r_0\right)+\pi ^2 (r_0-r)\right)\right],
\end{multline}
where $\mathcal{K}_2=\mathcal{K}_1\left(\eta ^2+r^2\right)$ and $\mathcal{K}_1$ is defined in Eq. \eqref{65}. The graphical behavior of EoS parameter $\omega_r$ for both models has been depicted in Fig. \ref{fig:5a}.\\
Now we consider another form of EoS, such as
\begin{equation}
P_t(r)=\omega_t(r)\rho(r),   
\end{equation}
where $\omega_t(r)$ is the EoS parameter which is a function of the radial coordinate $r$. With this form of EoS, we get the following differential equation
\begin{multline}
\alpha  \left[ b' \left(r \phi'+4 \omega +2\right)+b \left(r \phi'^2+\phi '+2 r \phi ''-\frac{2}{r}\right)\right.\\\left.
-r \left(\phi ' \left(r \phi '+2\right)+2 r \phi ''\right)\right]=0.
\end{multline}
Inserting the shape function \eqref{63} with the redshift function $\phi(r)=\frac{1}{2}\text{log}\left(1+\frac{\eta^2}{r^2}\right)$ in the last equation, we could obtain the tangential EoS parameter
\begin{multline}\label{1111}
\omega_t(r)=\frac{1}{4 \mathcal{K}_3}\left[\pi ^2 \left(5 \Lambda_i \lambda  \left(-\eta ^2 r^2 \left(r^3+8 r_0^3\right)+2 r^4\right.\right.\right.\\\left.\left.\left.
\times \left(r^3-4 r_0^3\right)-3 \eta ^4 r_0^3\right)+9 r^2 r_0^2 \left(-\eta ^2 r^2 (r+2 r_0)\right.\right.\right.\\\left.\left.\left.
+2 r^4 (r-2 r_0)-\eta ^4 r_0\right)\right)-\mathcal{K}_5\right],
\end{multline}
where
\begin{equation}
\mathcal{K}_3=\mathcal{K}_2 \left(\eta ^2+r^2\right),
\end{equation}
\begin{equation}
\mathcal{K}_4=\frac{r}{r_0}\left(\eta ^4+2 r^3 r_0+\eta ^2 r (4 r-r_0)\right),
\end{equation}
\begin{equation}
\mathcal{K}_5=\mathcal{D}_3 \mathcal{K}_4,
\end{equation}
and $\mathcal{D}_3$ is defined in Eq. \eqref{65a}.\\
Also, we could find the radial EoS parameter $\omega_r$ from the expression \eqref{64}
\begin{multline}\label{2222}
\omega_r(r)=\frac{1}{\mathcal{K}_6}\left[5 \pi ^2 \Lambda_i \lambda  r^2 \left(r_0^3-r^3\right)+9 r^4 r_0^2 \right.\\\left.
\times \left(720 \alpha  r_0 \left(\eta ^2+r r_0\right)+\pi ^2 (r_0-r)\right)\right],
\end{multline}
where $\mathcal{K}_6=3 \pi ^2 r_0^3 \left(\eta ^2+r^2\right) \left(5 \Lambda_i \lambda +3 r^2\right)$.\\
At throat, the expressions \eqref{1111} and \eqref{2222} reduce to
\begin{equation}\label{3333}
\omega_r\mid_{r=r_0}=\frac{2160 \alpha  r_0^4}{\pi ^2 \left(5 \Lambda_i \lambda +3 r_0^2\right)},
\end{equation}
\begin{equation}\label{4444}
\omega_t\mid_{r=r_0}=-\frac{\left(\eta ^2+2 r_0^2\right) \left(\pi ^2 \left(5 \Lambda_i \lambda +3 r_0^2\right)+2160 \alpha  r_0^4\right)}{4 \pi ^2 \left(\eta ^2+r_0^2\right) \left(5 \Lambda_i \lambda +3 r_0^2\right)}.
\end{equation}
It is evident that the right-hand side of Eq. \eqref{3333} is a positive quantity, whereas that of Eq. \eqref{4444} is a negative quantity. Thus it turns out that the radial EoS parameter $\omega_r$ increases and tangential $\omega_t$ decreases with the increase of the radial distance.
We have depicted the graphical behavior of the tangential EoS parameter in Fig. \ref{fig:8z}.

\section{Energy conditions}\label{sec5}
\begin{figure}[h]
    \centering
    \includegraphics[scale=0.5]{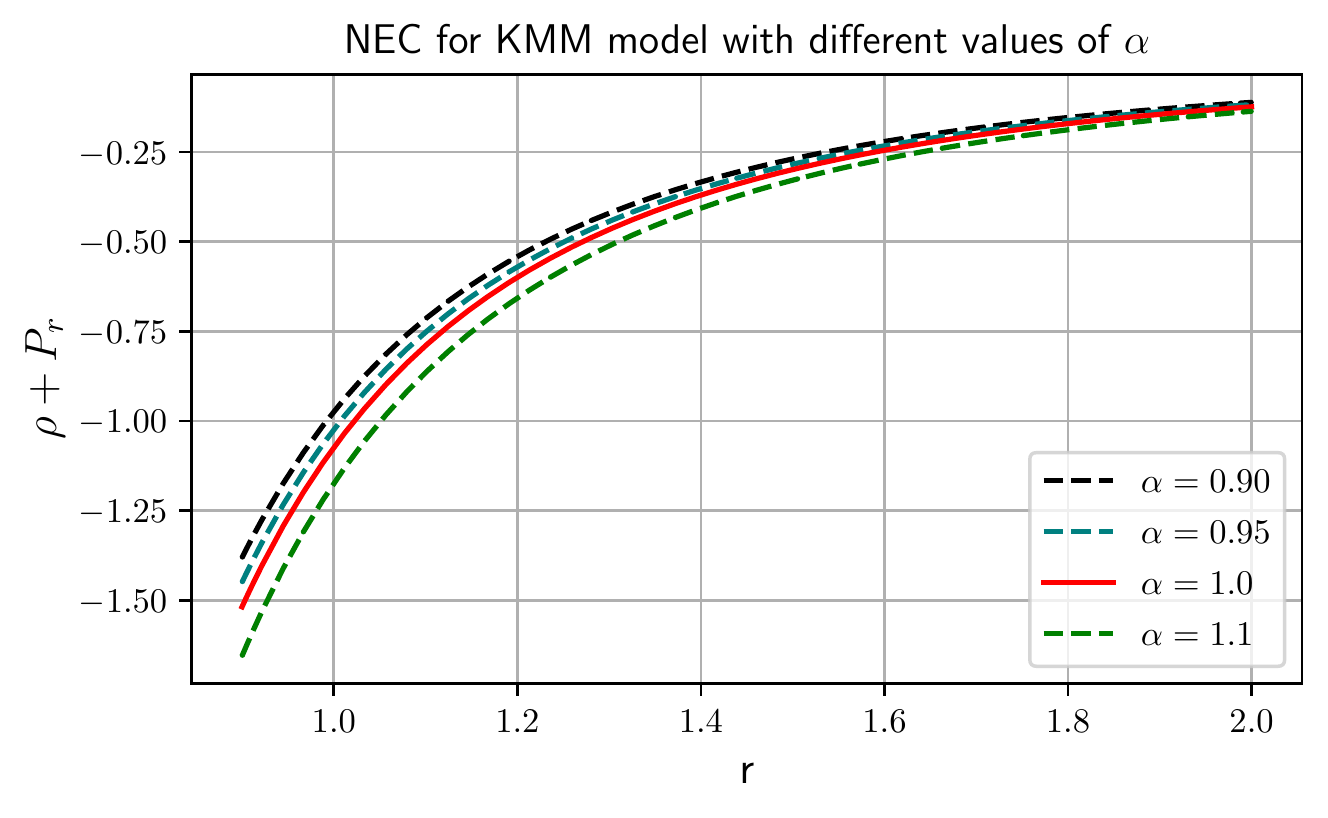}
    \includegraphics[scale=0.5]{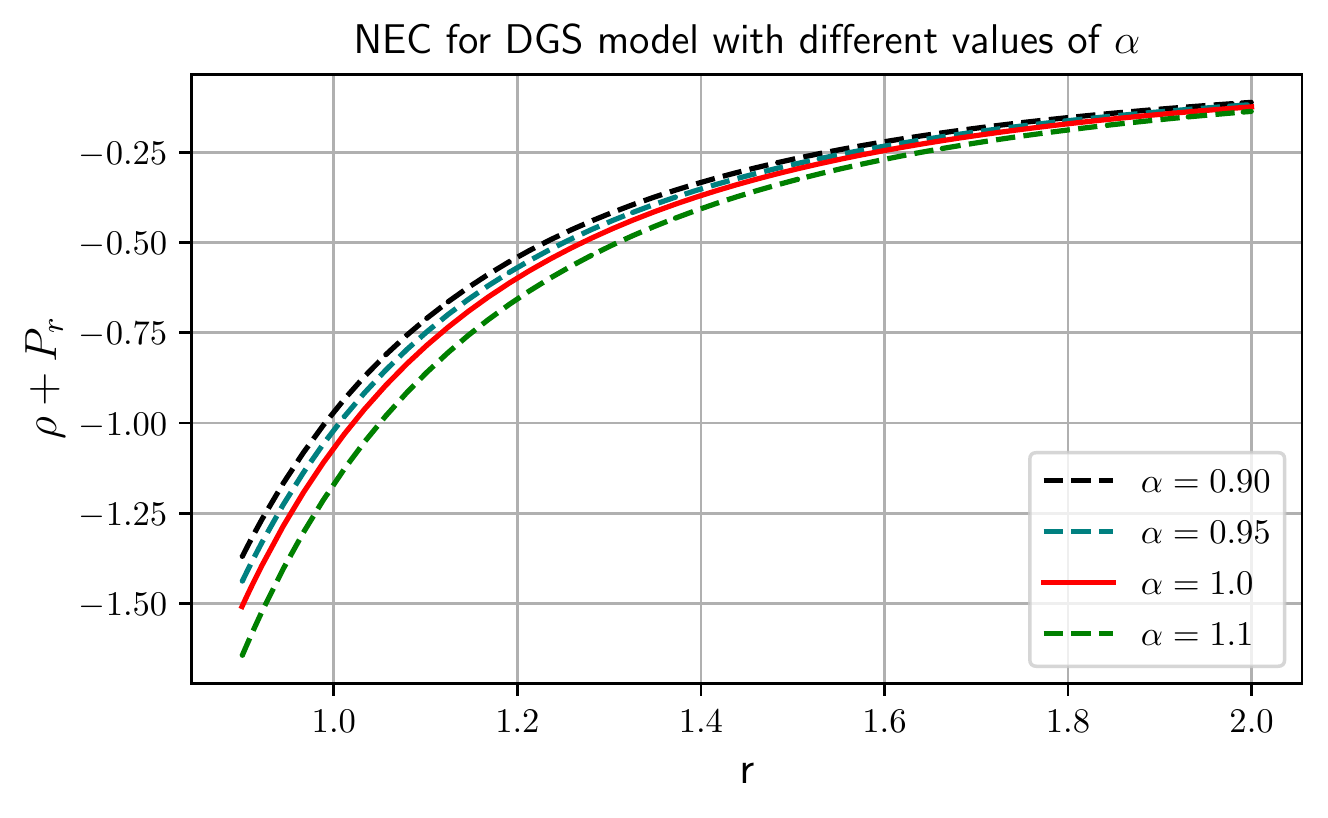}
    \caption{NEC for KMM and DGS models against radial distance $r$ using $\phi(r)=k$ with $r_0=1$ and GUP parameter $\lambda=0.1$. In the figure, $\alpha=1$ corresponds to the GR case.}
    \label{fig:5}
\end{figure}

\begin{figure}[h]
    \centering
    \includegraphics[scale=0.5]{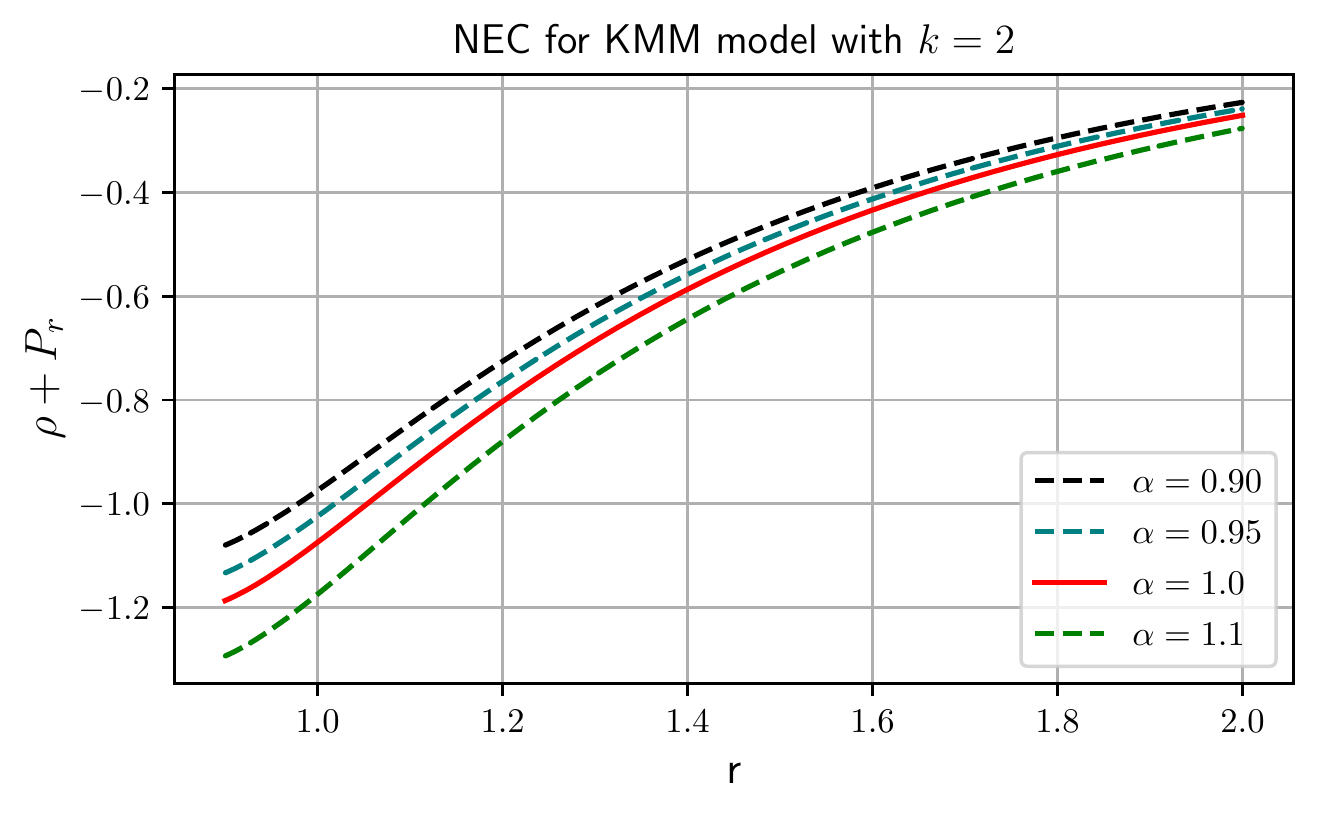}
    \caption{NEC for KMM model against radial distance $r$ using $\phi(r)=\frac{k}{r}$ with $r_0=1$ and GUP parameter $\lambda=0.1$. In the figure, $\alpha=1$ corresponds to the GR case.}
    \label{fig:6}
\end{figure}

\begin{figure}[h]
    \centering
    \includegraphics[scale=0.5]{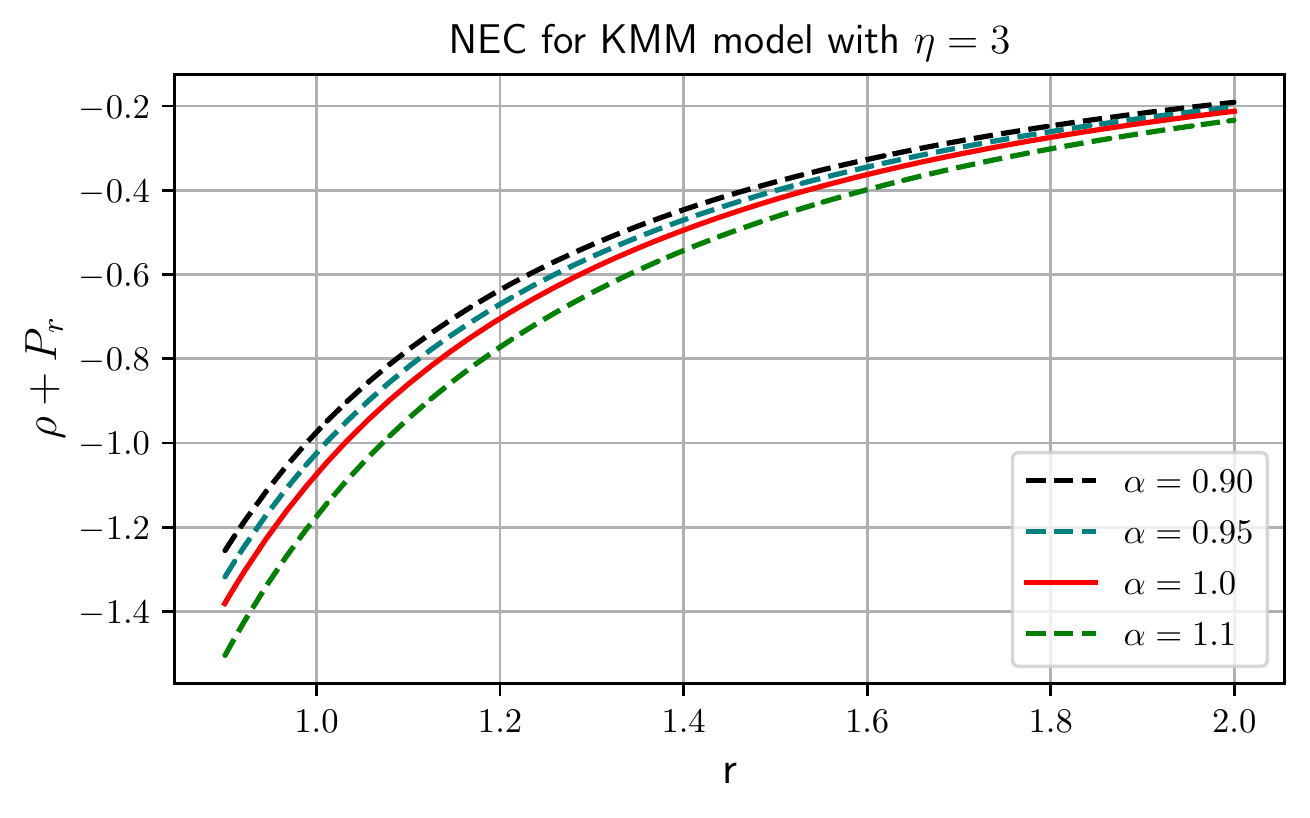}
    \caption{NEC for KMM model against radial distance $r$ using $\phi(r)=\frac{1}{2}\text{log}\left(1+\frac{\eta^2}{r^2}\right)$ with $r_0=1$ and GUP parameter $\lambda=0.1$. In the figure, $\alpha=1$ corresponds to the GR case.}
    \label{fig:7}
\end{figure}

In this section, we shall discuss the classical energy conditions developed from the Raychaudhuri equations. In GR, the wormhole solutions are maintained by exotic matter involving a stress-energy tensor that disrespects the NEC (indeed, it disobeys all the energy conditions \cite{Visser}). Note that the NEC can be defined as $$T_{\mu\nu}k^{\mu}k^{\nu}\geq 0,$$
where $k^{\mu}$ is the null vector. In this work, since we study with an anisotropic fluid of stress-energy tensor of the form \eqref{3b}, we have $\rho+P_i\geq 0$ where $i=r,\,t$.\\
Using Eqs. (\ref{7777}-\ref{9999}), the NEC for the GUP-corrected Casimir wormhole can be written from the above expression
\begin{multline}\label{71}
\rho+P_r=\frac{\alpha r_0}{r^3 \mathcal{D}_3}\left[r (r-r_0) \phi ' \mathcal{D}_2+\pi ^2 \left(5 \Lambda_i \lambda  \left(r^3-4 r_0^3\right)\right.\right.\\\left.\left.
+9 r^2 r_0^2 (r-2 r_0)\right)-\mathcal{D}_3\right],
\end{multline}
\begin{multline}\label{72}
\rho+P_t=\frac{\alpha r_0}{4 r^3 \mathcal{D}_3}\left[r \left(\phi ' \left(r (r-r_0) \phi ' \mathcal{D}_2+5 \pi ^2 \Lambda_i \lambda  \right.\right.\right.\\\left.\left.\left.
\times \left(r^3+2 r_0^3\right)+\mathcal{D}_4\right)+2 r (r-r_0) \phi '' \mathcal{D}_2\right)-2 \pi ^2 \right.\\\left.
\left(5 \Lambda_i \lambda  \left(r^3+2 r_0^3\right)+9 r^3 r_0^2\right)+\mathcal{D}_3\right],
\end{multline}
where, $\mathcal{D}_2$, $\mathcal{D}_3$ and $\mathcal{D}_4$ are already defined in Eqs. (\ref{65aa}-\ref{65aaa}).\\
Here, the GUP correction term is proportional to the uncertainty parameter $\lambda$. In the limit $\lambda \rightarrow 0$, the expressions \eqref{71} and \eqref{72} reduce to usual Casimir wormhole's NEC (see Eqs. (28) and (29) of Ref. \cite{Kazuharu}). One can notice that the right-hand side of Eq. \eqref{71} is a negative quantity for a radial distance $r \leq r_0$; hence, NEC is violated. Also, we observe that the contribution becomes more negative with the increase of GUP parameter $\lambda$ and model parameter $\alpha$.\\
At the throat of the wormhole, the above equations reduce to
\begin{equation}\label{73}
\rho+P_r\mid_{r=r_0}=-\left[\frac{\pi ^2 \left(5 \Lambda_i \lambda +3 r_0^2\right)}{2160 r_0^6}+\frac{\alpha }{r_0^2}\right],
\end{equation}
\begin{multline}\label{74}
\rho+P_t\mid_{r=r_0}=\frac{1}{8640 r_0^6}\left[\phi ' \left(\pi ^2 \left(5 \Lambda_i \lambda  r_0+3 r_0^3\right)\right.\right.\\\left.\left.
+2160 \alpha  r_0^5\right)-2 \pi ^2 \left(5 \Lambda_i \lambda +3 r_0^2\right)+4320 \alpha  r_0^4\right].
\end{multline}
It is transparent that the right-hand side of the Eq. \eqref{73} is a negative quantity for any positive $\alpha$. Thus, we could conclude that NEC is violated by the GUP-corrected Casimir wormhole at the throat. In Figs. \ref{fig:5}-\ref{fig:7}, we have plotted the graphs for NEC for both models with different redshift functions.

\section{GUP corrected Casimir wormholes for the Quadratic $f(Q)=Q+\gamma\,Q^2$ model}\label{sec5aa}
\begin{figure*}[h]
    \centering
    \includegraphics[scale=0.6]{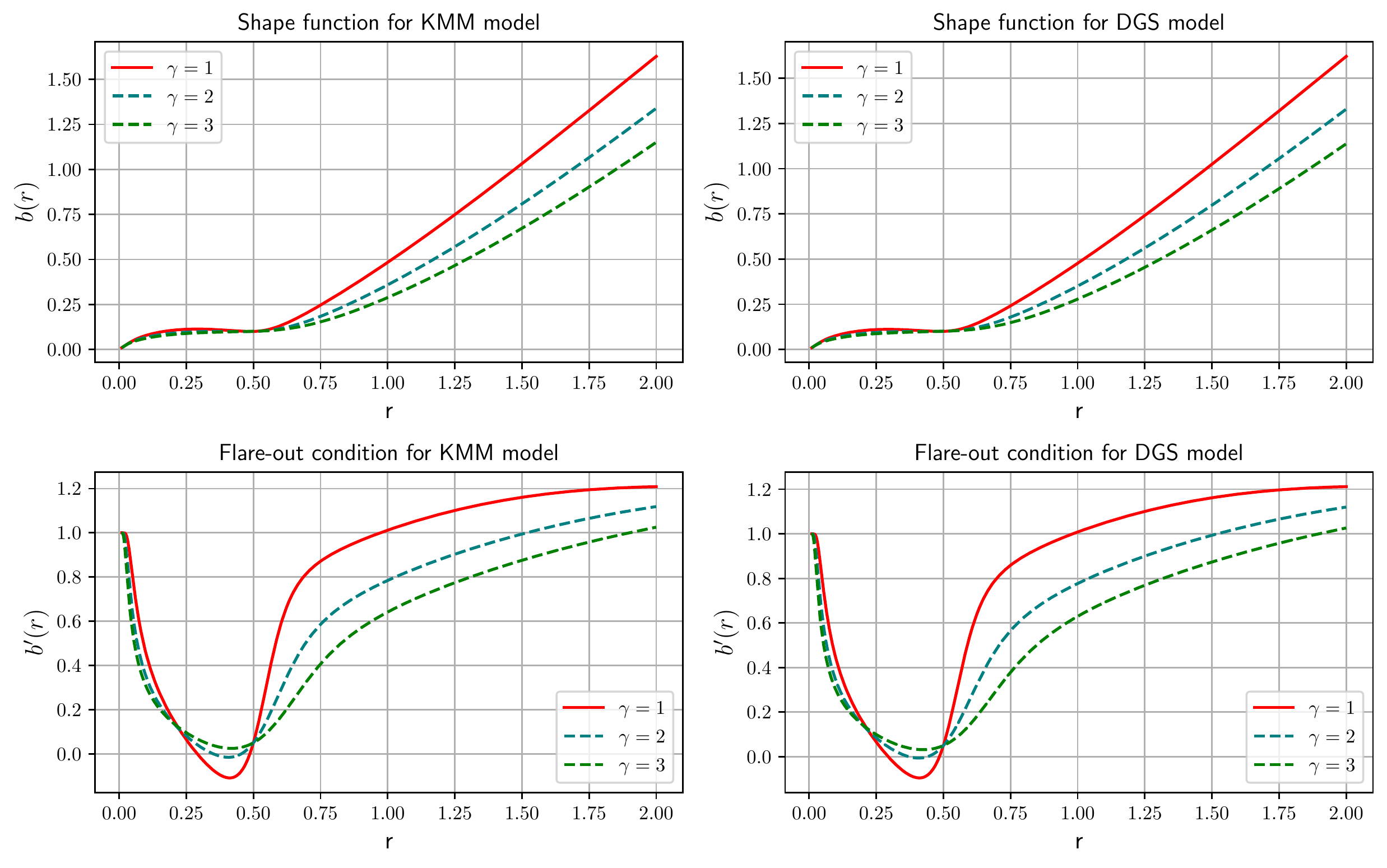}
    \caption{Shape function and Flare-out condition for KMM and DGS models for $\phi(r)=k$ under quadratic $f(Q)$ form. We fix the GUP parameter $\lambda=0.1$.}
    \label{fig:8a}
\end{figure*}

\begin{figure*}[h]
    \centering
    \includegraphics[scale=0.6]{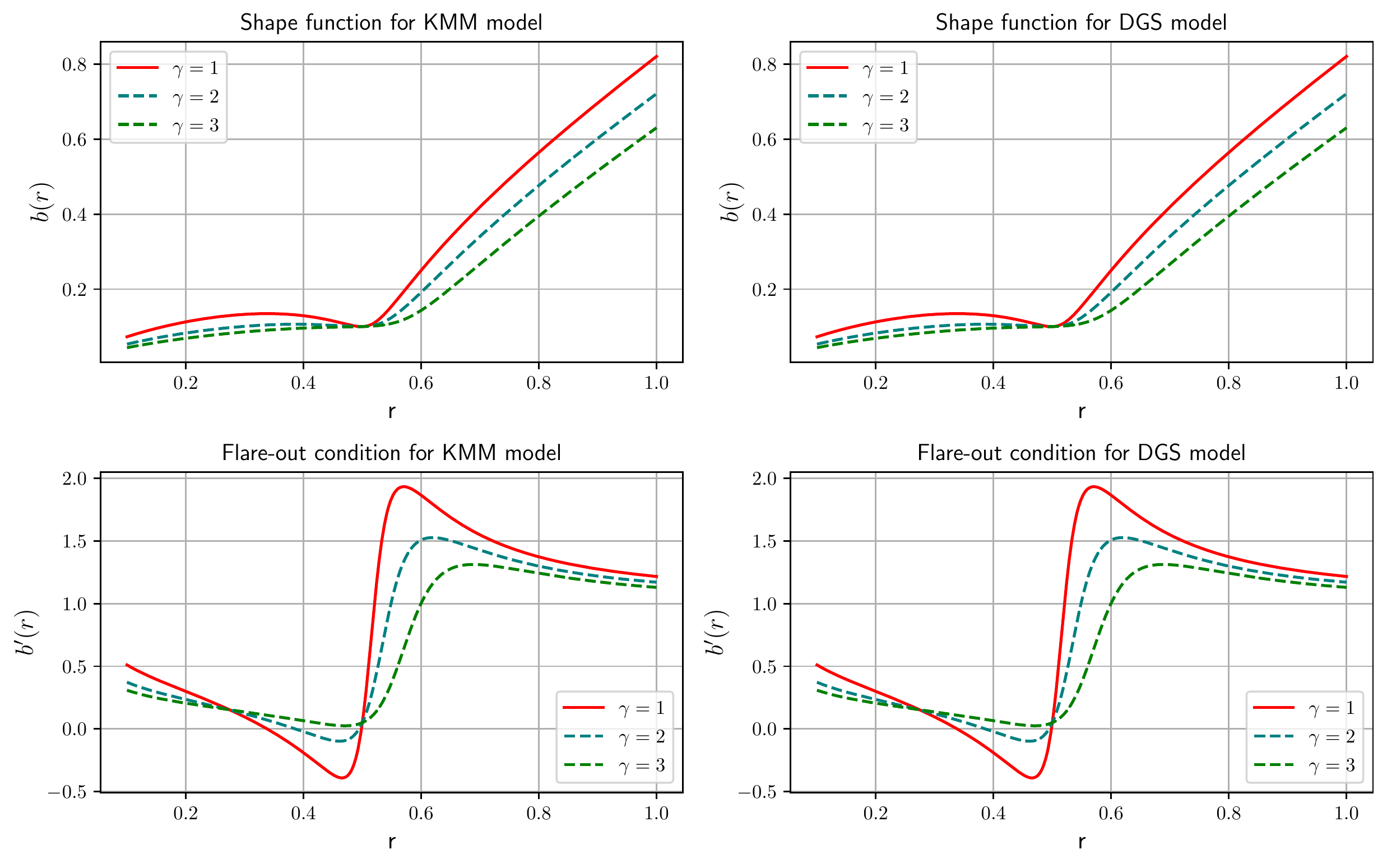}
    \caption{Shape function and Flare-out condition for KMM and DGS models for $\phi(r)=\frac{k}{r}$ under quadratic $f(Q)$ form. We fix the GUP parameter $\lambda=1$ and $k=1$.}
    \label{fig:8b}
\end{figure*}

\begin{figure}[h]
    \centering
    \includegraphics[scale=0.5]{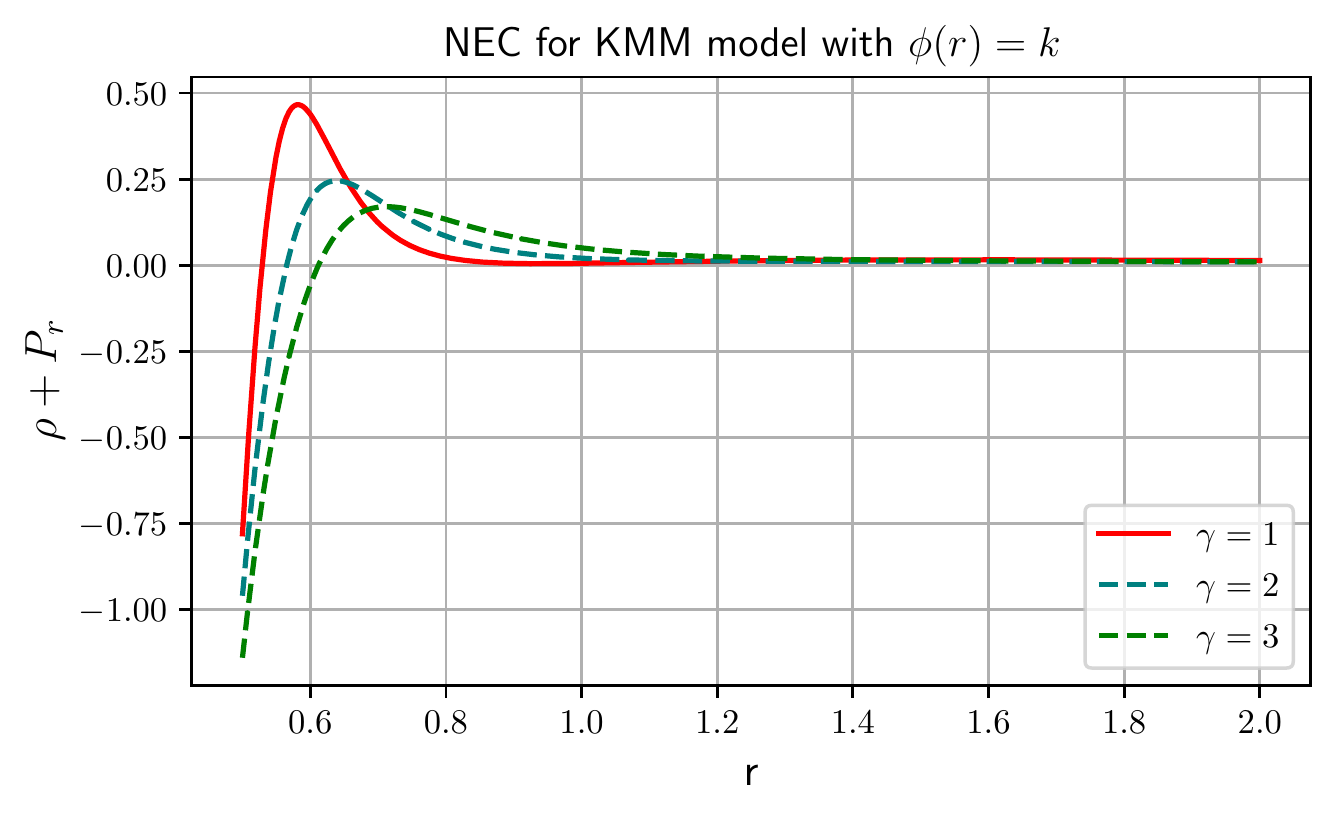}
    \includegraphics[scale=0.5]{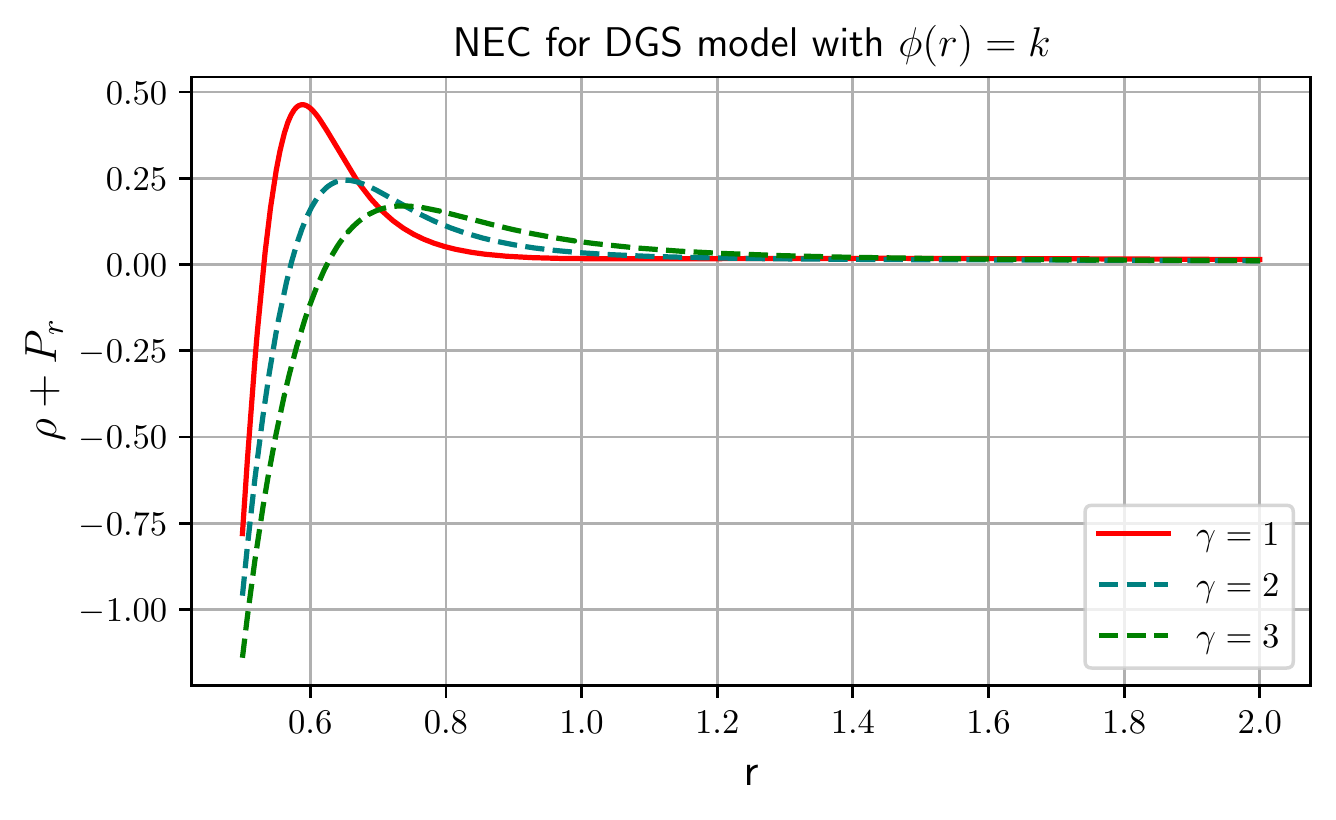}
    \caption{NEC for KMM and DGS models against radial distance $r$ for $f(Q)=Q+\gamma Q^2$ case with GUP parameter $\lambda=0.1$. In the figure, $\gamma=1$ corresponds to the GR case.}
    \label{fig:8c}
\end{figure}

\begin{figure}[h]
    \centering
    \includegraphics[scale=0.5]{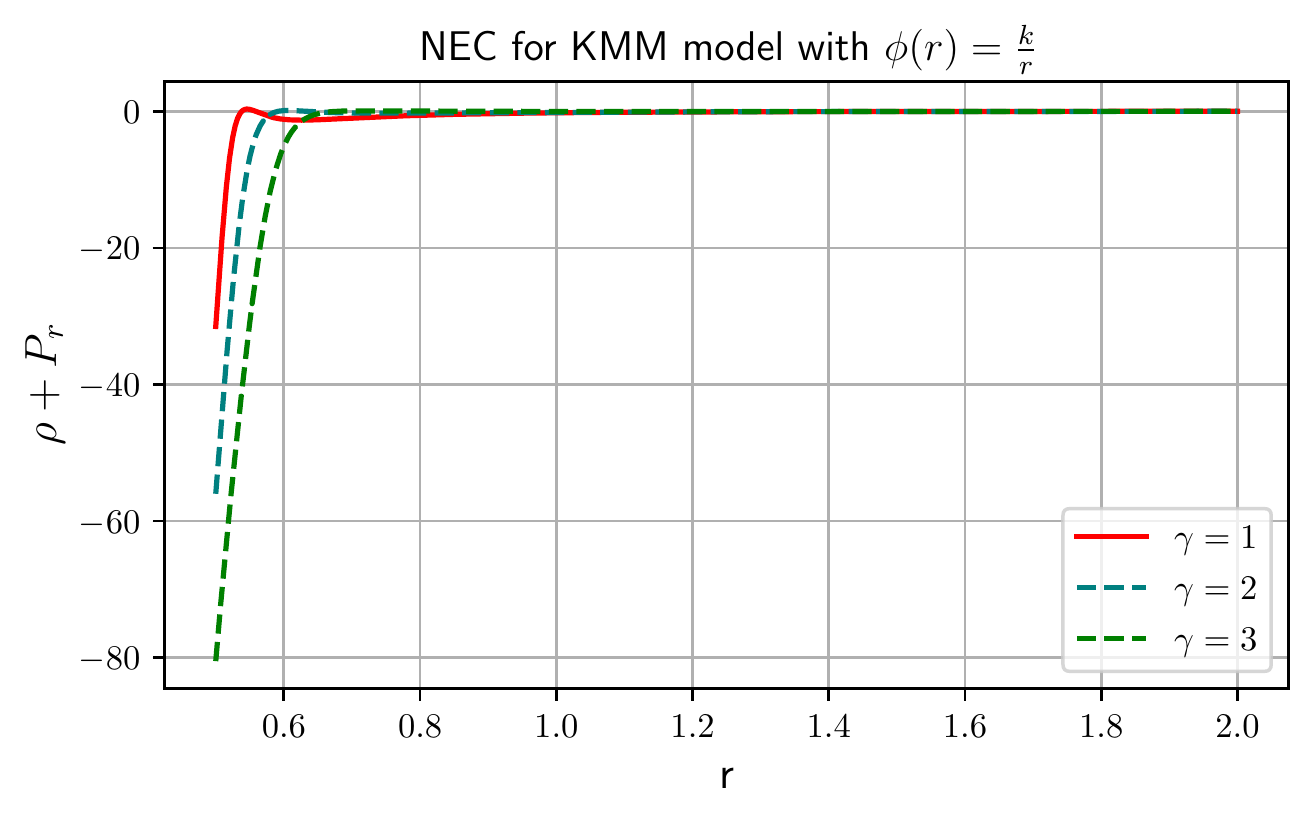}
    \includegraphics[scale=0.5]{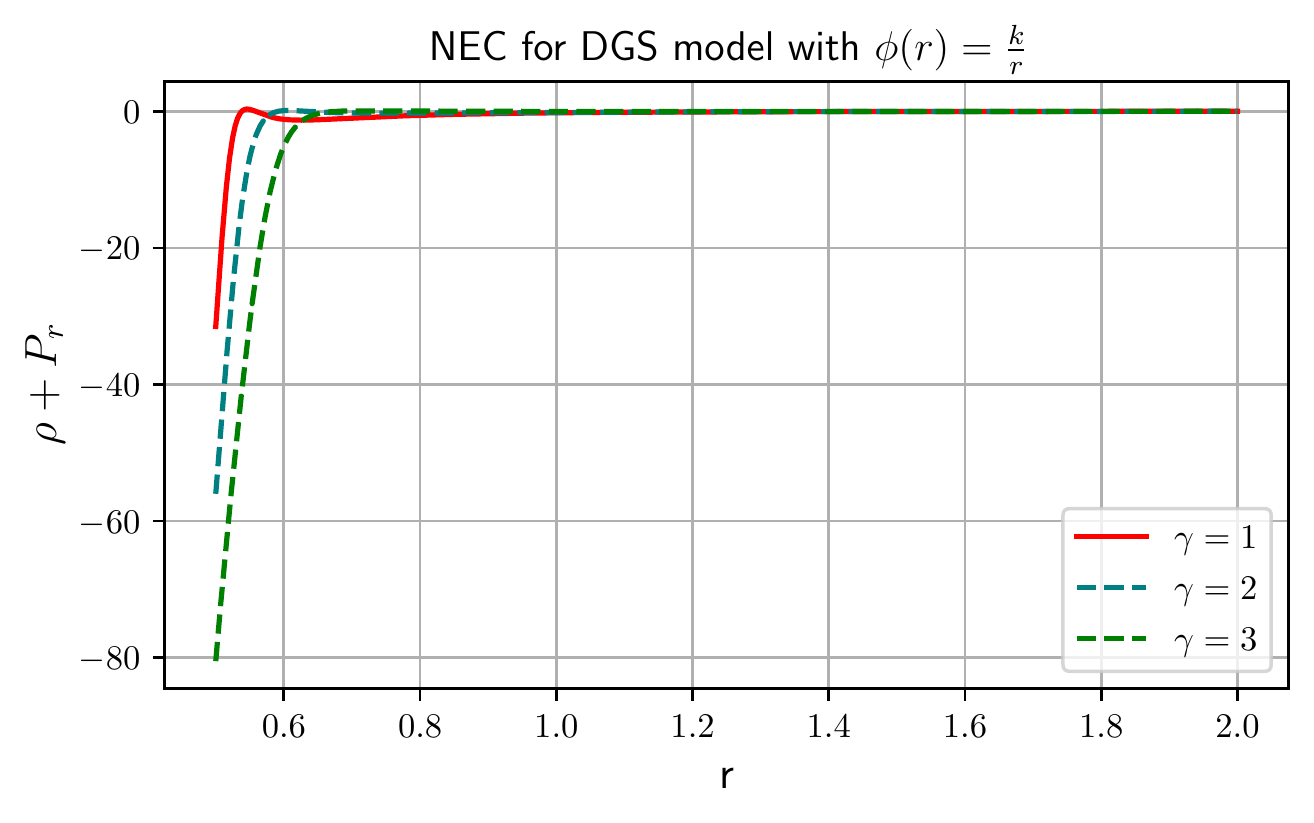}
    \caption{NEC for KMM and DGS models against radial distance $r$ for $f(Q)=Q+\gamma Q^2$ case with GUP parameter $\lambda=1$ and $k=1$. In the figure, $\gamma=1$ corresponds to the GR case.}
    \label{fig:8d}
\end{figure}

Here, we consider a quadratic form of $f(Q)$ model such as $f(Q)=Q+\gamma Q^2$ where $\gamma$ is the model parameter. One can note that, for $\gamma=0$, the above model will be equivalent to the GR case. This model has been used for stellar structure with polytropic EoS \cite{Zhai}. With the same model, Banerjee et al. \cite{Pradhan} discussed wormhole solutions for different shape functions. For the quadratic model, the generalized field equations (\ref{11}-\ref{13}) can be read as
\begin{multline}\label{14b}
\rho=\frac{1}{2 r^6 (r-b(r))^2}\left[r^2 b^2 \left(2 b' \left(12 \gamma +r^2+7 \gamma  r \phi'\right)\right.\right.\\\left.\left.
-\gamma  r \left(4 b''+\phi ' \left(r \phi'-8\right)+4 r \phi''\right)+3 \gamma  b'^2\right)-4 r^3 b b'\right.\\\left.
\times \left(2 \gamma  \left(b'(r)+r \phi'\right)+r^2\right)+2 r^6 b'+2 \gamma  r b^3 \left(r \left(2 b''\right.\right.\right.\\\left.\left.\left.
+\phi' \left(r \phi'-7\right)+4 r \phi''\right)-b' \left(3 r \phi'+7\right)-8\right)\right.\\\left.
+\gamma  b^4 \left(11-r \left(\phi' \left(r \phi'-6\right)+4 r \phi''\right)\right)\right],
\end{multline}
\begin{multline}\label{15b}
P_r=\frac{1}{2 r^6 (r-b)^2}\left[2 r^7 \phi'+r^2 b^2 \left(r \left(4 \left(\gamma  b''+\gamma  r \phi''+r\right)\right.\right.\right.\\\left.\left.\left.
+6 r^2 \phi'+13 \gamma  r \phi'^2\right)+2 \gamma  b' \left(r \phi'-8\right)+\gamma  b'^2\right)-2 r b^3 \right.\\\left.
\times \left(-6 \gamma +r \left(2 \gamma  \left(b''+2 r \phi''\right)+\left(\gamma +r^2\right) \phi '+7 \gamma  r \phi'^2\right)\right.\right.\\\left.\left.
+\gamma  b' \left(r \phi '-3\right)+r^2\right)-2 b \left(r^5 \left(2 \gamma  \phi'^2+3 r \phi '+1\right)\right.\right.\\\left.\left.
-2 \gamma  r^3 b'^2\right)+\gamma  b^4 \left(r \left(\phi ' \left(5 r \phi '+2\right)+4 r \phi ''\right)-7\right)\right],
\end{multline}
\begin{multline}\label{16b}
P_t=-\frac{1}{4 r^5 (r-b)^3}\left[\left(1-\frac{b}{r}\right) \left(-r \left(2 \gamma  b \left(-r b'+r\right.\right.\right.\right.\\\left.\left.\left.\left.
\times (b-r) \phi '+b\right)+r^3 (r-b)\right) \left(\left(b-r b'\right) \left((r-b) \phi '\right.\right.\right.\right.\\\left.\left.\left.\left.
+2\right)+r (r-b)^2 \phi '^2+2 (r-2 b) (r-b) \phi '+2 r (r-b)^2\right.\right.\right.\\\left.\left.\left.
\times \phi ''\right)+4 \gamma r (r-b)\phi ' \left(r^2 b \left(r \left(b''-2 \phi '+r \phi ''\right)-b'\right.\right.\right.\right.\\\left.\left.\left.\left.
\times \left(2 r \phi '+5\right) \right)+r^3 b' \left(b'+r \phi '\right)+r b^2 \left(-r \left(b''-4 \phi ' \right.\right.\right.\right.\right.\\\left.\left.\left.\left.\left.
+2 r \phi ''\right)+b'\left(r \phi '+3\right)+4\right)+b^3 \left(r^2 \phi ''-2 r \phi '-3\right)\right)\right.\right.\\\left.\left.
+2 b \left(-r b'+r(b-r)\phi '+b\right) \left(b \left(\gamma  \left(-r b'+r (b-r) \phi '\right.\right.\right.\right.\right.\\\left.\left.\left.\left.\left.
+b\right)-r^3\right)+r^4\right)\right)\right].
\end{multline}
A comparison of Eqs. \eqref{54aa} and \eqref{14b} yields the following non-linear differential equation:
\begin{multline}
\frac{1}{2 r^6 (r-b(r))^2}\left[r^2 b^2 \left(2 b' \left(12 \gamma +r^2+7 \gamma  r \phi'\right)-\gamma  r \right.\right.\\\left.\left.
\left(4 b''+\phi ' \left(r \phi'-8\right)+4 r \phi''\right)+3 \gamma  b'^2\right)-4 r^3 b b'\right.\\\left.
\times \left(2 \gamma  \left(b'(r)+r \phi'\right)+r^2\right)+2 r^6 b'+2 \gamma  r b^3 \left(r \left(2 b''\right.\right.\right.\\\left.\left.\left.
+\phi' \left(r \phi'-7\right)+4 r \phi''\right)-b' \left(3 r \phi'+7\right)-8\right)+\gamma  b^4 \right.\\\left.
\left(11-r \left(\phi' \left(r \phi'-6\right)+4 r \phi''\right)\right)\right]=-\frac{\pi^2}{720}\frac{1}{r^4}\\
\times \left[1+\frac{5}{3}\Lambda_i\left(\frac{\lambda}{r^2}\right)\right].
\end{multline}
whose analytic solution is also not possible. Thus, we numerically evaluate the shape function's possible form by solving the above equation.\\
Now, we shall examine the behavior of the shape functions acquired by the numerical technique and their corresponding essential properties for the existence of wormhole structures for the GUP-corrected Casimir energy density. For this purpose, we use  Mathematica numerical ODE solver \textit{NDSolve} with the initial conditions $b(0.5)=0.1$ and $b'(0.5)=0.05$. We have depicted the behavior of shape function and flaring out condition for different redshift functions in Figs. \ref{fig:8a} and \ref{fig:8b}. It can be observed that shape function $b(r)$ is showing increasing behavior in the entire space-time, but for increases in the value of the model parameter $\gamma$, it is decreasing monotonically. During the numerical plot, we noticed that the asymptotic flatness condition $\frac{b(r)}{r}$ is validated for a small radius, the reason being the non-linearity of the Lagrangian. It is known that the role of GUP is to correct the Casimir energy, and hence the non-linearity of the Lagrangian is inevitable due to quantum correction. Due to such small-scale quantum correction, we note that the asymptotic flatness condition might be satisfied far from the throat as the GUP approximation to Casimir energy is not valid far from the throat. Also, we located the wormhole throat at $r_0\approx 0.005$. Moreover, we checked the flaring out near the throat and found that very near the throat, it was satisfied. However, far from the throat flare-out condition will not be validated for both redshift functions.\\
Further, we have studied the energy conditions, especially NEC, near the wormhole throat, which are given in Figs. \ref{fig:8c} and \ref{fig:8d}. We observed that NEC is disrespected near the throat for both KMM and DGS models under both redshift functions. Also, violation of NEC becomes more if we increase the value of $\gamma$. However, NEC will be satisfied for large $r$, or far from the throat. Thus there exists a possibility of having a micro or tiny wormhole.

\section{ADM Mass of GUP Casimir wormhole}\label{sec6}
In order to give the physical meaning of $r_0$, we shall indeed show that $r_0=2M$ where $M$ is defined as the ADM mass for the metric \cite{adm0}. To find the ADM mass, we first write the formula for ADM mass
\begin{equation}\label{0909}
    M_{ADM}=\frac{1}{16\pi}\lim\limits_{r \to \infty}\sum_{\mu,\nu=1}^3\int_{S}(\partial_{\mu}h_{\mu \nu}-\partial_{\nu}h_{\mu \mu})N^{\nu}dS,
\end{equation}
where $h_{\mu\nu}$ is the induced metric over a constant $t$ slice, denoted by $\Sigma$, and $S$ is a topological two-sphere ($S^2$) embedded in the $\Sigma$, and $N^{\nu}$ is outward pointing unit normal over $S$, and $dS$ is the area element over the two-sphere \cite{adm1}.\\
Now our metric is given in Eq. \eqref{3a} is in a  4-dimensional space-time manifold, over a constant time slice $\Sigma$, the embedded metric takes the form
    \begin{equation}
       h_{\mu\nu}= ds^2_{\Sigma}=\left(1-\frac{b(r)}{r}\right)^{-1}dr^2+r^2d\theta^2+r^2\text{sin}^2\theta d\Phi^2.
    \end{equation}
Now, in general, the integral in the Eq \eqref{0909} is quite difficult, but for spherical symmetric metric, it is quite easy as it is done in \cite{adm2}.\\
The calculation is pretty involved as one has to switch from polar coordinate to Cartesian coordinates and use the symmetry property to get the answer. Here we mention the final solution for a spacetime metric is given by
\begin{equation}
    g_{ij}=\varphi dr^2 +\chi(r)r^2d\Omega^2.
\end{equation}
Also given that $\varphi$ and $\chi$ reach the asymptotic flat space-time limit as $\varphi-1=o(r^{-\frac{1}{2}})$, $\chi-1=o(r^{-\frac{1}{2}})$ and $\partial_r\varphi=o(r^{-\frac{3}{2}})$, $\partial_r\chi=o(r^{-\frac{3}{2}}).$
 Then $M_{ADM}$ is given by 
 \begin{equation}
    M_{ADM}= \lim\limits_{r \to \infty}\frac{1}{2}(-r^2\chi'+r(\varphi-\chi)).
 \end{equation}
For our case $\varphi(r)=\frac{1}{\left(1-\frac{b(r)}{r}\right)}$ and $\chi(r)=1$, so calculating the limit we get,
\begin{eqnarray}
     M_{ADM}&=&\lim\limits_{r \to \infty}\frac{1}{2}\left[-r^2\chi'+r(\varphi-\chi)\right]\\
     &=&\lim\limits_{r \to \infty}\frac{1}{2}r\left[\frac{1}{\left(1-\frac{b(r)}{r}\right)}-1\right]\\
     &=&\lim\limits_{r \to \infty}\frac{b(r)}{2}.
\end{eqnarray}
Here, we have used the asymptotic flatness condition $\lim\limits_{r \to \infty}\frac{b(r)}{r}=0$.
Using the shape function \eqref{63} in the last expression, we obtain
\begin{equation}\label{59}
    M_{ADM}=\frac{r_0}{2}-\frac{\xi_1}{10}\left(\frac{1}{r_0}\right)-\frac{\xi_1\lambda\Lambda_i}{18}\left(\frac{1}{r_0^3}\right).
\end{equation}
We know that to coincide with the Schwarzschild solution, $M=\frac{r_0}{2}$ should happen. Whereas the Eq. \eqref{59} clearly shows that under GUP, the ``effective mass" or ADM mass does change, and this is due to small-scale corrections that happen during consideration of GUP.

\section{Volume Integral Quantifier}\label{sec7}
In this section we shall investigate the amount of exotic matter necessary for maintaining a wormhole. Visser et al., \cite{Visser3} have proposed this VIQ technique to quantify the amount of average null energy condition (ANEC) violating matter present in space-time. The VIQ may be defined in terms of $\rho$ and radial pressure $P_r$ as 
\begin{equation}
Iv=\oint [\rho+P_r]dV
\end{equation}
where the volume can be read as $dV=r^2\,dr\,d\Omega$ with $d\Omega$ the solid angle. Since $\oint dV=2\int_{r_0}^{\infty}dV=8\pi \int_{r_0}^{\infty}r^2dr,$ we can write the last expression as follows
\begin{equation}
Iv=8\pi \int_{r_0}^{\infty}(\rho+P_r)r^2dr.
\end{equation}
It would be beneficial to have a wormhole whose field only varies from the throat $r_0$ to a particular radius $r_1$ with $r_1\geq r_0,$ and then we can have
\begin{equation}\label{1122}
Iv=8\pi \int_{r_0}^{r_1}(\rho+P_r)r^2dr.
\end{equation}
with the help of Eqs. \eqref{7777} and \eqref{8888}, integrating the above expression for the redshift function $\phi(r)=k,$ we obtain
\begin{multline}
Iv=\frac{1}{3\mathcal{F}_1}\left[3 r_1^3 \log \frac{r_1}{r_0} \left(\pi ^2 \left(5 \Lambda_i \lambda +9 r_0^2\right)-6480 \alpha  r_0^4\right)\right.\\\left.
+2 \pi ^2 \left(10 \Lambda_i \lambda  \left(r_0^3-r_1^3\right)+27 r_1^2 r_0^2 (r_0-r_1)\right)\right],
\end{multline}
where $$\mathcal{F}_1=6480 r_1^3 r_0^3.$$
Also, for $\phi(r)=\frac{k}{r},$ we obtain from the integral \eqref{1122}
\begin{multline}
Iv=\frac{1}{12 r_1 r_0 \mathcal{F}_1}\left[(r_1-r_0) \left(54 r_1^2 r_0^2 \left(1440 r_1 \alpha  k r_0^2 \right.\right.\right.\\\left.\left.\left.
+\pi ^2 (k r_0-r_1 (k+4 r_0))\right)-5 \pi ^2 \Lambda_i \lambda  \left(r_1^3 (9 k+16 r_0)\right.\right.\right.\\\left.\left.\left.
+r_1^2 r_0 (16 r_0-3 k)+r_1 r_0^2 (16 r_0-3 k)-3 k r_0^3\right)\right)\right.\\\left.
+12 r_1^4 r_0 \log \frac{r_1}{r_0} \left(9 r_0^2 \left(\pi ^2-720 \alpha  r_0 (k+r_0)\right)+5 \pi ^2 \Lambda_i \lambda \right)\right].
\end{multline}
Moreover, we could find the volume integral $Iv$ for the redshift function $\phi(r)=\frac{1}{2}\text{log}\left(1+\frac{\eta^2}{r^2}\right)$ as
\begin{multline}
Iv=\frac{1}{2\mathcal{F}_1 \eta ^3}\left[r_1^3 \left(\mathcal{F}_2 \eta ^3  \left(\log \frac{r_1^2+\eta ^2}{\eta ^2+r_0^2}\right)+\mathcal{F}_3\right.\right.\\\left.\left.
\times \left(6480 \alpha  \eta ^4+\pi ^2 \left(9 \eta ^2-5 \Lambda_i \lambda \right)\right)\right)-2 \pi ^2 \eta  (r_1-r_0) \right.\\\left.
\times \left(5 \Lambda_i \lambda  \left(\eta ^2 \left(r_1^2+r_1 r_0+r_0^2\right)+r_1^2 r_0^2\right)+9 r_1^2 \eta ^2 r_0^2\right)\right],
\end{multline}
where $$\mathcal{F}_2=\pi ^2 \left(5 \Lambda_i \lambda +9 r_0^2\right)-6480 \alpha  r_0^4\,,$$
$$\mathcal{F}_3=2 r_0^3\left[ \tan ^{-1}\left(\frac{r_0}{\eta }\right)- \tan ^{-1}\left(\frac{r_1}{\eta }\right)\right]\,.$$
\begin{figure}[h]
    \centering
    \includegraphics[scale=0.63]{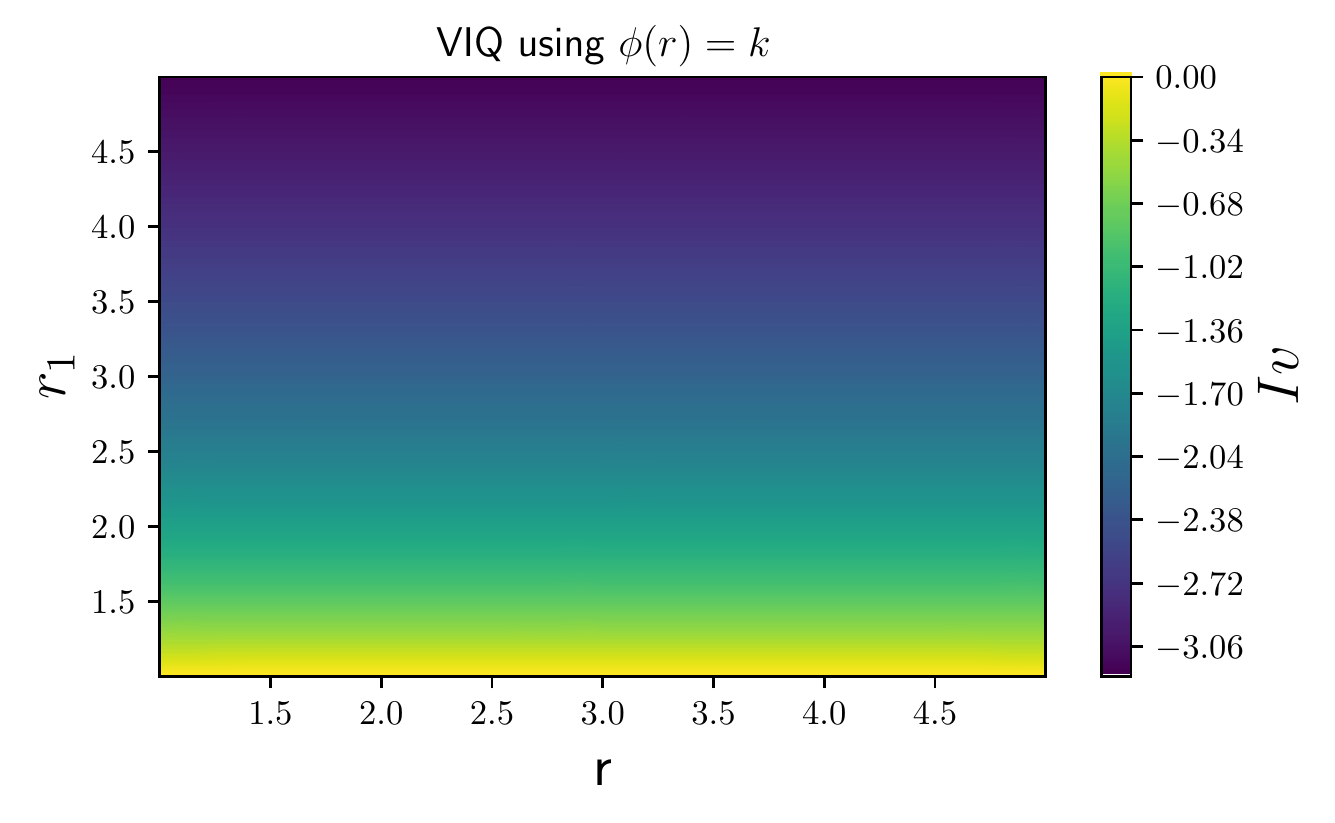}
    \caption{The evolution of $Iv$ against $r$ and $r_1$ for KMM model. We consider GUP parameter $\lambda=0.1$, $r_0=1$ and $\alpha=0.95$.}
    \label{fig:8}
\end{figure}
\begin{figure}[h]
    \centering
    \includegraphics[scale=0.63]{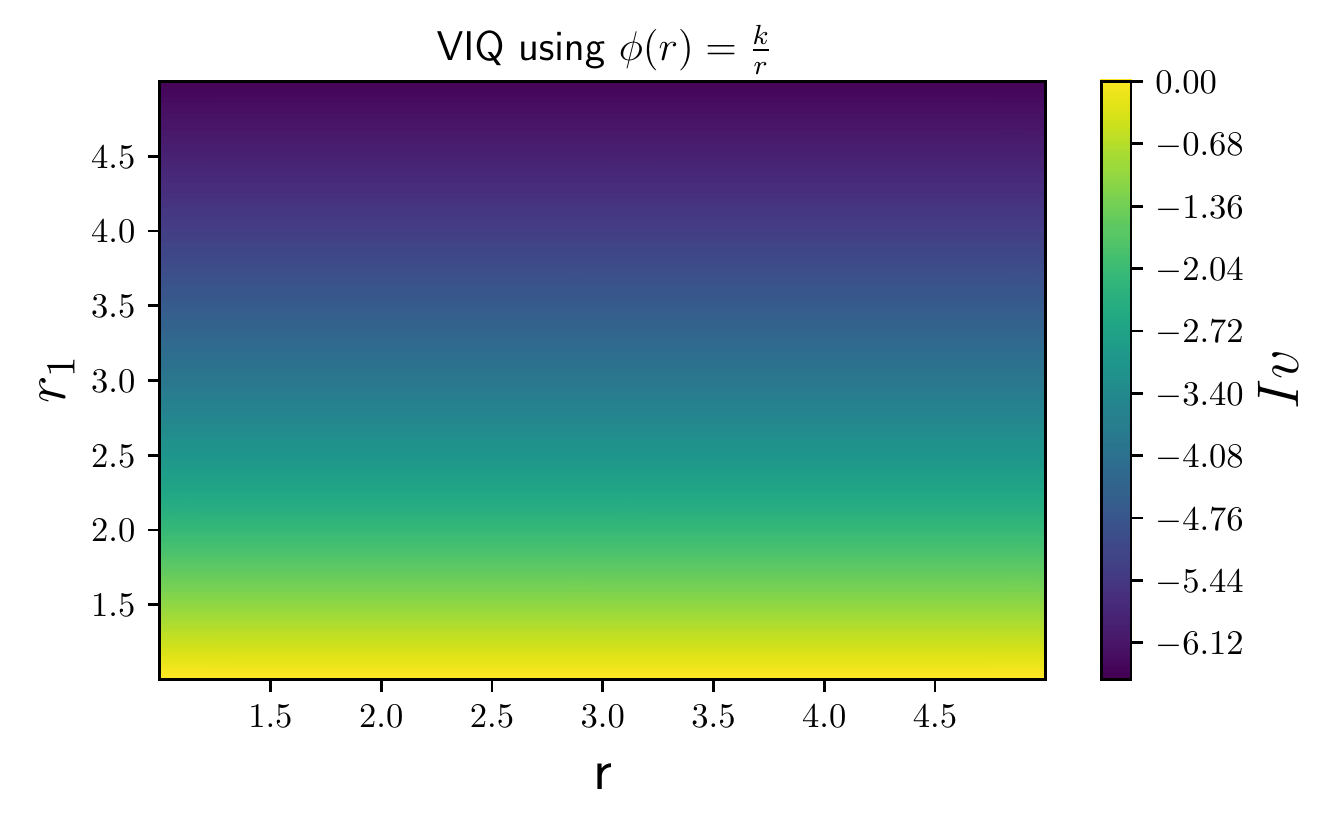}
    \caption{The evolution of $Iv$ against $r$ and $r_1$ for DGS model. We consider GUP parameter $\lambda=0.1$, $r_0=1$, $k=2$ and $\alpha=0.95$.}
    \label{fig:9}
\end{figure}
\begin{figure}[h]
    \centering
    \includegraphics[scale=0.63]{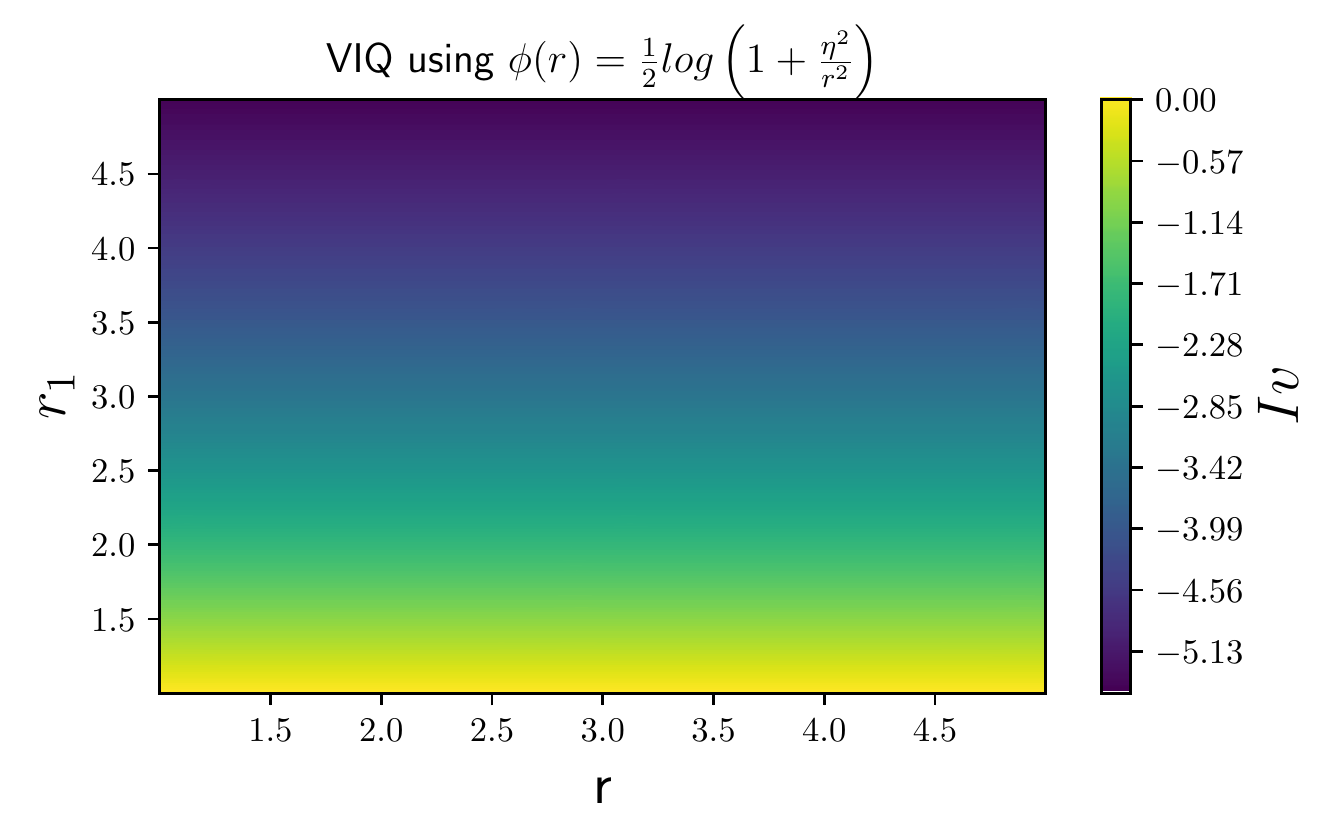}
    \caption{The evolution of $Iv$ against $r$ and $r_1$ for KMM model. We consider GUP parameter $\lambda=0.1$, $r_0=1$, $\eta=3$ and $\alpha=0.95$.}
    \label{fig:10}
\end{figure}
In Figs. (\ref{fig:8}-\ref{fig:10}), we have depicted the nature of volume integral $Iv$. Note that for $r_1\rightarrow r_0$, we should find $Iv\rightarrow 0$. One may observe from the figures that our obtained solutions satisfy the condition. Thus this reveals the existence of spacetime geometries containing traversable wormholes sustained by arbitrarily small amounts of exotic matter. In fact, the total amount of ANEC-violating matter can be reduced by considering suitable wormhole geometry. Readers may check the Refs. \cite{Baransky,Channuie1} for a detailed discussion on this interesting topic.
\section{Conclusions}\label{sec8}
In this work, we have investigated the effect of the Generalized Uncertainty Principle (GUP) on the Casimir wormhole space-time in modified symmetric teleparallel gravity. The Casimir effect that occurs, attributable to the distorted quantized field of the vacuum between two parallel plane plates, is associated with exotic energy and pressure, which may be possible in the laboratory. Such types of exotic matter disrespect the energy conditions. Since in GR, wormhole material content must be exotic, and it should disobey some energy conditions and even present a negative mass. Hence the quantum nature of the Casimir effect might help model these exotic objects. Here, we studied the exact analytic solutions of Morris-Throne wormhole field equations for $f(Q)$ gravity, describing the Casimir wormhole with the effect of GUP correction. However, GUP is not sufficient as a fundamental probe to the minimal length scale needed for quantum gravity as discussed in \cite{rainbowsmolin,sabinegup}, but it is a way to probe the length scales where the quantum gravity effects get nonnegligible. Also, much work has been studied on the field-theoretic aspects of GUP, like showing how the ultraviolet divergence behaves. By calculating the first-order loop diagrams, \cite{regula}, it has been shown that the field theory is renormalizable and that there is no ultraviolet divergence.\\
In this paper, we have employed two GUP relations: the KMM model and the DGS model, to show the effect of GUP. In \cite{Channuie1}, the authors have discussed the effect of GUP in Casimir wormholes by invoking the above GUP relations in GR. Later Tripathy \cite{Tripathy} has investigated the GUP effect in $f(R,T)$ gravity. Here, we have utilized the mentioned GUP models to check the effect of GUP in the Casimir wormhole in this recently modified gravity. For this study, we have considered two $f(Q)$ models, such as linear $f(Q)=\alpha Q+\beta$ and quadratic $f(Q)=Q+\gamma\,Q^2$ models, where $\alpha$, $\beta$ and $\gamma$ are model parameters. Also, we have considered three different constant and non-constant redshift functions to acquire asymptotically flat wormhole solutions under GUP-affected Casimir density. In order to obtain the EoS parameters $\omega_r(r)$ and $\omega_t(r)$, we did use two famous EoS relations defined by $P_r=\omega_r(r)\rho$ and $P_t=\omega_t(r)\rho$, respectively. Our main theoretical observations are discussed below.\\
For the linear model, we have compared the GUP-corrected Casimir energy density with the energy density of the modified gravity and integrated it to obtain the shape function of the wormhole space-time metric. The resulting shape function respects the flare-out condition under asymptotic background. Graphically we have shown the effect of the GUP parameter and modified gravity in shape functions. One may notice an increase in the GUP parameter $\lambda$ when the radial distance far from the throat decreases the shape function, whereas an increase in the model parameter $\alpha$ results in an increase in it. Nevertheless, in the throat, this effect is not substantial.\\
We have also investigated the behavior of EoS parameters for radial and tangential pressure to the Casimir energy density under different redshift functions. We have observed that the radial EoS parameter increases with the increase of radial distance $r$ and suitable parameters, whereas the tangential EoS parameter shows the opposite behavior. We can see the effect of modified gravity in the EoS parameter to a large extent, at least at distances away from the throat.\\
Again, we have studied the energy conditions, especially NEC, at the wormhole's throat with radius $r_0$ for both models. For each redshift function, we have noticed that NEC is violated in a small neighborhood of the throat. The violation contribution becomes more negative for an increase in $\alpha$. Thus this has demonstrated that some arbitrary amount of small quantity disrespects the classical energy condition at the wormhole's throat.\\
Further, for the quadratic model, We have used numerical techniques by setting some initial conditions and studying the graphical behavior of shape functions and energy conditions. We noticed that the shape function showed positively decreasing behavior as the values of $\gamma$  increased for both KMM and DGS models. Also, the flaring-out condition is satisfied near the throat, whereas, for large $r$, this condition will no longer be validated. Moreover, we have investigated the energy conditions and confirmed that NEC is violated for both models near the throat. As stated earlier, Banerjee et al. \cite{Pradhan} discussed wormhole solutions for different shape functions and confirmed that solutions might not exist for considered shape functions under this quadratic model. But, from this study, it is worth mentioning that wormhole solutions could be possible numerically using appropriate initial conditions. However, this analysis shows the possibility of the existence of a macro or tiny wormhole.
\\
Furthermore, we have examined the ADM mass of the wormhole and VIQ to study the amount of exotic matter required at the throat for a traversable wormhole. Our analysis found that a small amount of exotic matter is necessary for a traversable wormhole.\\
Recently, the authors of \cite{Kazuharu} have studied Casimir wormholes without GUP in $f(Q)$ gravity. They have considered three different systems, such as parallel plate, cylindrical plate, and two spheres, and investigated the effect of these systems in wormhole geometry. Here, in this work, we have extended the above work by correcting the GUP effect in the parallel plate and systematically investigated the impact of GUP in Casimir wormholes in this modified symmetric teleparallel gravity.\\
Moreover, one may study the effect of GUP Casimir wormhole in other modified gravity, such as in $f(T)$ gravity, as both $f(T)$ and $f(Q)$ models are indistinguishable at the cosmological background level \cite{Koivisto}. Further, one can calculate the corrections of Casimir energy up to the next leading order using GUP corrected QED (Quantum electrodynamics), as is done in \cite{qedgup1,qedgup2} and can explore the significance in wormhole solutions.\\
\section*{Acknowledgments}
ZH acknowledges the Department of Science and Technology (DST), Government of India, New Delhi, for awarding a Senior Research Fellowship (File No. DST/INSPIRE Fellowship/2019/IF190911). SG acknowledges the Council of Scientific and Industrial Research (CSIR), Government of India, New Delhi, for junior research fellowship (File no.09/1026(13105)/2022-EMR-I). PKS acknowledges National Board for Higher Mathematics (NBHM) under the Department of Atomic Energy (DAE), Govt. of India for financial support to carry out the Research project No.: 02011/3/2022 NBHM(R.P.)/R\&D II/2152 Dt.14.02.2022. We are very much grateful to the honorable referees and to the editor for the illuminating suggestions that have significantly improved our work in terms
of research quality, and presentation.


\begin{thebibliography}{52}
\footnotesize
\bibitem{abbott1} B. P. Abbott, et al., LIGO Scientific Collaboration, Virgo Collaboration, \textit{Phys. Rev. Lett.} \textbf{121}, 129902 (2018).
\bibitem{Abbott2} B. P. Abbott, et al., LIGO Scientific Collaboration, Virgo Collaboration, \textit{Phys. Rev. Lett.} \textbf{116}, 061102 (2016).
\bibitem{Akiyama} K. Akiyama, et al., Event horizon telescope, \textit{Astrophys. J.} \textbf{875}, L1 (2019).
\bibitem{Khatsymovsky} V. Khatsymovsky, \textit{Phys. Lett. B} \textbf{320}, 234 (1994).
\bibitem{Halilsoy} A. Ovgun, M. Halilsoy, \textit{Astrophys. Space Sci.} \textbf{361}, 214 (2016).
\bibitem{Ray} F. Rahaman, P. K. F. Kuhfittig, S. Ray, N. Islam, \textit{Eur. Phys. J. C} \textbf{74}, 2750 (2014).
\bibitem{Bambi} C. Bambi and D. Stojkovic, \textit{Universe} \textbf{7}, 136 (2021).
\bibitem{Flamm} L. Flamm, \textit{Phys. Z.} \textbf{17}, 448 (1916).
\bibitem{Rosen} A. Einstein, N. Rosen, \textit{Ann. Phys.} \textbf{2}, 242 (1935).
\bibitem{Misner} C. W. Misner and J. A. Wheeler, \textit{Annals Phys.} \textbf{2}, 525 (1957).
\bibitem{Kruskal} M. D. Kruskal, \textit{Phys. Rev.} \textbf{119}, 1743-1745 (1960).
\bibitem{Fuller} R. W. Fuller and J. A. Wheeler,  \textit{Phys. Rev.} \textbf{128}, 919-929 (1962).
\bibitem{Eardley} D. M. Eardley, \textit{Phys. Rev. Lett.} \textbf{33}, 442-444 (1974).
\bibitem{Thorne/1988} M. S. Morris, K. S. Thorne, \textit{Am. J. Phys.} \textbf{56}, 395-412 (1988).
\bibitem{Visser} M. Visser, Lorentzian Wormholes: From Einstein to Hawking, first ed., American Institute of Physics, New York,
(1996).
\bibitem{Lobo/2005} F. S. N. Lobo, \textit{Phys. Rev. D} \textbf{71}, 084011 (2005).
\bibitem{Sushkov} S. V. Sushkov, \textit{Phys. Rev. D} \textbf{71}, 043520 (2005).
\bibitem{Parsaei} F. S. N. Lobo, F. Parsaei, and N. Riazi, \textit{Phys. Rev. D} \textbf{87}, 084030 (2013).
\bibitem{Moradpour11} Y. Heydarzade, N. Riazi, and H. Moradpour, \textit{Can. J. Phys.} \textbf{93}, 1523 (2015).
\bibitem{Anabalon} A. Anabalon and A. Cisterna, \textit{Phys. Rev. D} \textbf{85}, 084035 (2012).
\bibitem{Kuhfittig} M. Jamil, P. K. F. Kuhfittig, F. Rahaman and S. A. Rakib, \textit{Eur. Phys. J. C} \textbf{67} 513 (2010).
\bibitem{Balakin} A. B. Balakin, J. P. S. Lemos and A. E. Zayats, \textit{Phys. Rev. D} \textbf{81}, 084015 (2010).
\bibitem{Hansen} J. Hansen, D. I. Hwang and D. H. Yeom, \textit{J. High Energy Phys.} \textbf{0911}, 016 (2009).
\bibitem{Diaz} P. F. Gonzalez-Diaz, \textit{Phys. Rev. D} \textbf{68}, 084016 (2003).
\bibitem{Dehghani} M. H. Dehghani and S. H. Hendi, \textit{Gen. Rel. Grav.} \textbf{41}, 1853 (2009).
\bibitem{Visser1} M. Visser, \textit{Phys. Rev. D} \textbf{39}, 3182 (1989).
\bibitem{Visser2} M. Visser, \textit{Nuclear Phys. B} \textbf{328}, 203 (1989).
\bibitem{Visser3} M. Visser, S. Kar and N. Dadhich, \textit{Phys. Rev. Lett.} \textbf{90}, 201102 (2003).
\bibitem{Duplessis} Francis Duplessis and Damien A. Easson, \textit{Phys. Rev. D} \textbf{92}, 043516 (2015).
\bibitem{Simeone} M. G. Richarte, C. Simeone, \textit{Phys. Rev. D} \textbf{80}, 104033 (2009).
\bibitem{Aguirre} E. F. Eiroa, G.F. Aguirre, \textit{Eur. Phys. J. C} \textbf{72}, 2240 (2012).
\bibitem{Shaikh} R. Shaikh \textit{Phys. Rev. D} \textbf{98}, 064033 (2018).
\bibitem{Garcia} N. M. Garcia and F. S. N. Lobo, \textit{Phys. Rev. D} \textbf{82}, 104018 (2010).
\bibitem{Garcia1} N. M. Garcia and F. S. N. Lobo, \textit{Classical Quantum Gravity} \textbf{28}, 085018 (2011).
\bibitem{Bronnikov1} K. A. Bronnikov and A. M. Galiakhmetov, \textit{Grav. Cosmol} \textbf{21}, 283 (2015).
\bibitem{Bronnikov2} K. A. Bronnikov and A. M. Galiakhmetov, \textit{Phys. Rev. D} \textbf{94}, 124006 (2016).
\bibitem{Mehdizadeh1} M. R. Mehdizadeh and A. H. Ziaie, \textit{Phys. Rev. D} \textbf{95}, 064049 (2017).
\bibitem{Moradpour} H. Moradpour, N. Sadeghnezhad, and S. H. Hendi, \textit{Can. J. Phys.} \textbf{95}, 1257 (2017).
\bibitem{ad1} A. Malik, et al., \textit{Eur. Phys. J C} \textbf{83}, 522 (2023). 
\bibitem{ad2} A. Malik, et al., \textit{New Astron.} \textbf{104}, 102071 (2023). 
\bibitem{ad3} Z. Asghar, et al., \textit{Chin. J. Phys.} \textbf{83}, 427 (2023). 
\bibitem{ad4} M. F. Shamir, et al., \textit{Fortschr. Phys.} \textbf{70}, 2200134 (2022). 
\bibitem{ad5} M. F. Shamir, et al., \textit{Chin. J. Phys.} \textbf{73}, 634-648 (2021). 
\bibitem{ad6} M. F. Shamir, et al., \textit{Int. J. Mod. Phys. A} \textbf{36}, 2150021 (2021). 
\bibitem{ad7} M. F. Shamir and A. Malik, \textit{Chin. J. Phys.} \textbf{69}, 312 (2021).
\bibitem{Tsukamoto} N. Tsukamoto, T. Harada, K. Yajima, \textit{Phys. Rev. D} \textbf{86}, 104062 (2012).
\bibitem{Shaikh3} R. Shaikh, \textit{Phys. Rev. D} \textbf{98}, 024044 (2018).
\bibitem{Nedkova} G. Gyulchev, P. Nedkova, V. Tinchev, S. Yazadjiev, \textit{Eur. Phys. J. C} \textbf{78}, 544 (2018).
\bibitem{Jamil1} S. Bahamonde, M. Jamil, P. Pavlovic, M. Sossich, \textit{Phys. Rev. D} \textbf{94}, 044041 (2016).
\bibitem{Jamil2} S. Bahamonde, U. Camci, S. Capozziello, M. Jamil, \textit{Phys. Rev. D} \textbf{94}, 084042 (2016).
\bibitem{Fayyaz} M. F. Shamir and I. Fayyaz, \textit{Eur. Phys. J. C} \textbf{80}, 1102 (2020).
\bibitem{Horvat11} A. DeBenedictis and D. Horvat, \textit{Gen Relativ Gravit} \textbf{44}, 2711-2744 (2012).
\bibitem{Eiroa} E. F. Eiroa and G. F. Aguirre, \textit{Eur. Phys. J. C} \textbf{76}, 132 (2016).
\bibitem{Karakasis} T. Karakasis, E. Papantonopoulos, C. Vlachos, \textit{Phys. Rev. D} \textbf{105}, 024006 (2022).
\bibitem{Sossich} S. Bahamonde, M. Jamil, P. Pavlovic, M. Sossich, \textit{Phys. Rev. D} \textbf{94}, 044041 (2016).
\bibitem{Sahoo} P. H. R. S. Moraes and P. K. Sahoo, \textit{Phys. Rev. D} \textbf{96}, 044038 (2017).
\bibitem{Ahmad} M. Zubair, S. Waheed, Y. Ahmad, \textit{Eur. Phys. J. C} \textbf{76}, 444 (2016).
\bibitem{Ilyas} Z. Yousaf, M. Ilyas and M. Zaeem-ul-Haq Bhatti, \textit{Eur. Phys. J. Plus} \textbf{132}, 268 (2017).
\bibitem{Khurshudyan} E. Elizalde and M. Khurshudyan, \textit{Phys. Rev. D} \textbf{98}, 123525 (2018).
\bibitem{Harko/2012} C. G. B\"ohmer, T. Harko, and F. S. N. Lobo, \textit{Phys. Rev. D} \textbf{85} 044033 (2012).
\bibitem{Jamil/2013} M. Jamil, D. Momeni and  R. Myrzakulov, \textit{Eur. Phys. J. C} \textbf{73}, 2267 (2013).
\bibitem{Shamaila} M. Sharif and S. Rani, \textit{Phys. Rev. D} \textbf{88}, 123501 (2013).
\bibitem{Ali} Ali \"Ovg\"un, \textit{Phys. Rev. D} \textbf{98}, 044033 (2018).
\bibitem{Nawazish} M. Sharif, I. Nawazish, S. Hussain, \textit{Eur. Phys. J. C} \textbf{80} 783 (2020).
\bibitem{Godani} N. Godani, S. Debata, S. K. Biswal, G. C. Samanta, \textit{Eur. Phys. J. C}, \textbf{80}, 40 (2020).
\bibitem{2} R. Myrzakulov, L. Sebastiani, S. Vagnozzi, S. Zerbini, \textit{Class. Quantum Grav.} \textbf{33}, 125005 (2016).
\bibitem{Tayde} M. Tayde, Z. Hassan, P. K. Sahoo, S. Gutti, \textit{Chin. Phys. C}, \textbf{46}, 115101 (2022).
\bibitem{Nazir} M. Sharif and Kanwal Nazir, \textit{Ann. Phys.} \textbf{393}, 145-166 (2018).
\bibitem{Nazir1} I. Fayyaz and M. F. Shamir, \textit{Eur. Phys. J. C} \textbf{80}, 430 (2020).
\bibitem{Laurentis} S. Capozziello and M. De Laurentis, \textit{Phys. Rept.} \textbf{509}, 167 (2011).
\bibitem{Krssak} M. Krssak, R. J. van den Hoogen, J. G. Pereira, C. G. B\"ohmer, and A. A. Coley, \textit{Class. Quant. Grav.} \textbf{36}, 183001 (2019).
\bibitem{Nester} James M. Nester and  Hwei-Jang Yo, \textit{Chin.J.Phys.} \textbf{37}, 113 (1999).
\bibitem{Kalay} M. Adak, M. Kalay, and O. Sert, \textit{Int. J. Mod. Phys.} \textbf{D15}, 619 (2006).
\bibitem{Conroy} A. Conroy and T. Koivisto, \textit{Eur. Phys. J. C} \textbf{78}, 923 (2018).
\bibitem{Jimenez} J. B. Jimenez, L. Heisenberg, and T. S. Koivisto, \textit{Phys. Rev. D} \textbf{98}, 044048 (2018).
\bibitem{Soudi} I. Soudi, G. Farrugia, V. Gakis, J. L. Said, E. N. Saridakis, \textit{Phys. Rev. D} \textbf{100}, 044008 (2019).
\bibitem{Banos} R. Lazkoz, F. S. N. Lobo, M. O. Banos, V. Salzano, \textit{Phys. Rev. D} \textbf{100}, 104027 (2019).
\bibitem{Salzano} I. Ayuso, R. Lazkoz, V. Salzano, \textit{Phys. Rev. D} \textbf{103}, 063505 (2021).
\bibitem{Koivisto11} B. J. Barros, T. Barreiro, T. Koivisto, N. J. Nunes, \textit{Phys. Dark Univ.} \textbf{30}, 100616 (2020).
\bibitem{Anagnostopoulos} F. K. Anagnostopoulos, S. Basilakos, E. N. Saridakis, \textit{Phys. Lett. B} \textbf{822}, 136634 (2021).
\bibitem{Fell} F. D'Ambrosio, S. D. B. Fell, L. Heisenberg, S. Kuhn \textit{Phys. Rev. D} \textbf{105}, 024042 (2022).
\bibitem{Zhai} Rui-Hui Lin and Xiang-Hua Zhai, \textit{Phys. Rev. D} \textbf{103}, 124001 (2021).
\bibitem{Wang2} W. Wang, H. Chen, T. Katsuragawa, \textit{Phys. Rev. D} \textbf{105},  024060 (2022).
\bibitem{Hassan1} Zinnat Hassan, Sanjay Mandal, P.K. Sahoo, \textit{Forts. Phys.} \textbf{69}, 2100023 (2021).
\bibitem{Mustafa} G. Mustafa, Z. Hassan, P.H.R.S. Moraes, P.K. Sahoo, \textit{Phys. Lett. B} \textbf{821}, 136612 (2021).
\bibitem{1} M. Calza and L. Sebastiani, \textit{arXiv}, arXiv:2208.13033 [gr-qc].
\bibitem{Sokoliuk} O. Sokoliuk, Z. Hassan, P.K. Sahoo, A. Baransky, \textit{Ann. Phys.} \textbf{443}, 168968 (2022).
\bibitem{Sahoo111} G. Mustafa, Z. Hassan, P. K. Sahoo, \textit{Ann. Phys.} \textbf{437}, 168751 (2022).
\bibitem{Hassan3} S. Mandal, G. Mustafa, Z. Hassan, P. K. Sahoo \textit{Phys. Dark Univ.} \textbf{35}, 100934 (2022).
\bibitem{casimir}  H. B. G. Casimir,  D. Polder, \textit{Phys. Rev.} \textbf{73}, 360-372 (1948).
\bibitem{lifshitz} I. E. Dzyaloshinskii, E. M. Lifshitz, Lev P Pitaevskii,  \textit{Soviet Physics Uspekhi.}, \textbf{4}, 153 (1961).
\bibitem{experiment} S. K. Lamoreaux,  \textit{ Phys. Rev. Lett.}, \textbf{78}, 5-8 (1997).
\bibitem{bressi} G. Bressi, G. Carugno, R. Onofrio, G. Ruoso \textit{Phys. Rev. Lett.}, \textbf{88}, 041804 (2002).
\bibitem{Garattini} R. Garattini, \textit{Eur. Phys. J. C} \textbf{79}, 951 (2019).
\bibitem{rainbowsmolin}  J. Magueijo, L. Smolin \textit{Class. Quantum Grav.} \textbf{21} 1725 (2004).
\bibitem{sabinegup} S. Hossenfelder 
\textit{Living Rev. Relativity},\textbf{16}, 2, (2013).
\bibitem{ellis} A. Camelia et al, \textit{Nature} \textbf{393}, 763-765, (1998).
\bibitem{dasexp} S. Das, E. C. Vagenas, \textit{Phys. Rev. Lett.} \textbf{101} 221301 (2008).
\bibitem{gupmain} M. Bishop , J. Contreras, D. Singleton , \textit{universe} \textbf{8} 192, (2022).
\bibitem{gupcasimir} A. M. Frassino, O. Panella, \textit{Phys.Rev.D}  \textbf{85} 045030 (2012).
\bibitem{3} M. Faizal, A. F. Ali, S Das, \textit{Int. J. Mod. Phys. A} \textbf{32}, 1750049 (2017).
\bibitem{4} I Arraut, D Batic, M Nowakowski, \textit{Class. Quantum Grav.} \textbf{26}, 125006 (2009).
\bibitem{7} R. Garattini, \textit{Eur. Phys. J. C} \textbf{79}, 951 (2019).
\bibitem{Channuie} D. Samart, T. Tangphati, P. Channuie, \textit{Nuclear Phys. B} \textbf{980}, 115848 (2022).
\bibitem{zee} A. Zee, Quantum Field Theory in a Nutshell, second ed, Princeton University Press, (2010).
\bibitem{paddy} T. Padmanabhan, Quantum Field Theory The Why, What and How, first ed, Springer, (2016).
\bibitem{Bezerra} G. Alencar, V. B. Bezerra, C. R. Muniz, \textit{Eur. Phys. J. C} \textbf{81}, 924 (2021).
\bibitem{Fuenmayor} R. Avalos, E. Fuenmayor, E. Contreras, \textit{Eur. Phys. J. C} \textbf{82}, 420, (2022).
\bibitem{Khabibullin} A. R. Khabibullin $et\,al.$ \textit{Class. Quantum Grav.} \textbf{23}, 627 (2006).
\bibitem{kmm} A. Kempf, G Mangano, R. B. Mann \textit{Phys.Rev.D} \textbf{52}, 1108, (1995).
\bibitem{gupmath} A. Kempf, 
  \textit{J. Math. Phys.} \textbf{35}, 4483 (1994).
\bibitem{gupmath2} A. Kempf, \textit{arxiv}, arXiv:hep-th/9810215.
\bibitem{dgs} S. Detournay, C. Gabriel, P. Spindel, \textit{Phys.Rev.D} \textbf{66} 125004 (2002).
\bibitem{regula} A. Kempf, G. Mangano  \textit{Phys.Rev.D} \textbf{55} 7909 (1997).
\bibitem{Solanki} R. Solanki, A. De, P. K. Sahoo, \textit{Phys. Dark Univ.} \textbf{36}, 101053 (2022).
\bibitem{Parbati} P. Sahoo, A. De, Tee-How Loo, P. K. Sahoo, \textit{arXiv}, arXiv:2110.11768
\bibitem{Kazuharu} Zinnat Hassan, Sayantan Ghosh, P. K. Sahoo, Kazuharu Bamba, \textit{Eur. Phys. J. C}, \textbf{82}, 1116 (2022).
\bibitem{Pradhan} A. Banerjee, A. Pradhan, T. Tangphati, F. Rahaman, \textit{Eur. Phys. J. C} \textbf{81}, 1031 (2021).
\bibitem{adm0} R. M. Wald, \textit{General Relativity} (University of Chicago Press, Chicago, 1984).
\bibitem{adm1} R. Shaikh, \textit{Phys.Rev.D} \textbf{98}, 024044 (2018).
\bibitem{Channuie1} K. Jusufi, P. Channuie, M. Jamil, \textit{Eur. Phys. J. C} \textbf{80}, 127 (2020).
\bibitem{adm2} P. T. Chrusciel, Lectures on Energy in General Relativity, Krakow, 2010.
\bibitem{Baransky} O. Sokoliuk, S. Mandal, P. K. Sahoo, A. Baransky, \textit{Eur. Phys. J. C} \textbf{82}, 280 (2022).
\bibitem{Tripathy} S. K. Tripathy, \textit{Phys. Dark Univ.} \textbf{31}, 100757 (2021).
\bibitem{Koivisto} J. B. Jimenez, L. Heisenberg, T. Koivisto, and S. Pekar, \textit{Phys. Rev. D} \textbf{101}, 103507 (2020).
\bibitem{qedgup1} O. Panella \textit{Phys.Rev.D} \textbf{76}, 045012 (2007).
\bibitem{qedgup2} K. Nouicer, \textit{J. Phys. A} \textbf{38}, 10027 (2005).

\end{thebibliography}
\end{document}